\documentstyle[12pt,aaspp4,epsf]{article}

\def\ltsim{ \,{}^<_\sim\, }
\def\gtsim{ \,{}^>_\sim\, }
\def\etal{{\it et~al.}}
\def\ie{i.e.}
\def\eg{e.g.}
\def\cf{cf.}

\lefthead{Harris \etal }
\righthead{M87 Globular Clusters}

\begin{document}

\title{M87, Globular Clusters, and Galactic Winds:
Issues in Giant Galaxy Formation}

\author{William E.~Harris\altaffilmark{1}}
\affil{Department of Physics \& Astronomy, 
McMaster University, Hamilton, Ontario L8S 4M1
\\Electronic mail: harris@physics.mcmaster.ca}

\author{Gretchen L.~H.~Harris\altaffilmark{1}}
\affil{Department of Physics, University of Waterloo,
Waterloo, Ontario N2L 3G1
\\Electronic mail: glharris@astro.uwaterloo.ca}

\author{Dean E.~McLaughlin\altaffilmark{1,2}}
\affil{Department of Physics \& Astronomy, 
McMaster University, Hamilton, Ontario L8S 4M1
\\Electronic mail: dean@crabneb.berkeley.edu}

\altaffiltext{1}{Visiting Astronomer, Canada-France-Hawaii Telescope, operated
 by
the National Research Council of Canada, le Centre National de la Recherche
Scientifique de France, and the University of Hawaii.}

\altaffiltext{2}{Hubble Fellow; now at Department of Astronomy, University of California,
Berkeley CA 94720}
 
\clearpage

\begin{abstract}

We have used the High Resolution Camera at the Canada-France-Hawaii
Telescope to obtain $VRI$ photometry of the globular clusters in
the innermost $140''$ of the M87 halo.   The results are used to
discuss several issues concerning the formation and evolution of
globular cluster systems in supergiant ellipticals like M87.
Our principal results are as follows:  (1) From our deep 
$R-$band photometry of the cluster population, 
we find {\it no significant change} in the globular cluster luminosity
function (GCLF) with galactocentric radius, 
for cluster masses $M \gtsim 10^5 M_{\odot}$. 
This result places constraints on current theoretical predictions 
of the rate of cluster evolution by tidal shocking and
evaporation, indicating that the main effects of dynamical
evolution may be only on lower-mass clusters ($\ltsim 10^5 M_{\odot}$)
that are below the faint limit of most current observations.
(2) Combining our $(V-I)$ color indices with other data in the literature,
we derive the metallicity gradient and mean metallicity 
of the clusters from $r = 9''$ out to $r \simeq 500''$ ($\sim 0.7$ to 35 kpc).
Within the core radius $r_c = 1'$ of the globular cluster
system, the metallicity distribution is uniform, but at larger radii
the mean metallicity declines steadily as $Z/Z_{\odot} \sim r^{-0.9}$.
(3) The various options for explaining the existence of high specific
frequency galaxies like M87 are evaluated.  We argue that 
alternatives involving secondary evolution
(such as the formation of many globular clusters during mergers, or the
existence of a population of intergalactic globular clusters centered
on the same location as the central elliptical) are capable of
modifying the characteristics of the GCS in distinctive ways, but are not
likely to be the primary explanation for high$-S_N$ galaxies.
(4) We offer a new explanation for the large observed $S_N$ range
among brightest cluster ellipticals (BCGs).  It is suggested that 
these central supergiant E galaxies formed in an exceptionally
turbulent or high-density environment which favored a very rapid initial
star formation burst.  As a result, a much higher than average
fraction of the protogalactic gas was driven out in a galactic wind
not long after the first globular clusters were formed, 
thus artificially boosting the specific frequency of the remaining galaxy.
From a total sample of 30 BCGs, we derive empirical scaling relations
which relate to this hypothesis.  Our analysis favors the view that BCGs
began forming at redshifts $z \gtsim 5$, distinctly earlier than
most other galaxies.

\end{abstract}

\keywords{Stellar Systems:  Globular clusters; Normal galaxies}

\clearpage

\section{INTRODUCTION}

Although much is now known about the characteristics of globular cluster
systems (GCSs) in large galaxies (\eg, \cite{har91}), the properties of
globular clusters in the {\it inner} regions of these galaxies are still
poorly understood, for the simple reason that such objects are extremely
difficult to observe against the strong background light of their host galaxy.
The innermost few kiloparsecs are, however, precisely the regions where the
strongest traces of dynamical evolution of the clusters should be seen, and
are thus of considerable interest as tests of the many dynamical models now in
the literature (see, \eg, \cite{oka95}; \cite{cpr96}; \cite{gne97}; 
\cite{ost97}; \cite{ves97} for
comprehensive recent discussions).  Do the inner clusters exhibit the expected
history of disruption through their spatial distribution, luminosity
distribution (GCLF), orbital distribution function, or local specific
frequency?  What fraction of the globular clusters that might have initially
formed in these inner regions survived successfully to the present day?  At present
there are few galaxies for which relevant GCS data exist to constrain the
models strongly.

From the theoretical side, an additional major problem is to decide which of
the systemic properties of the GCS that we see today (spatial distribution,
GCLF, and so on) are due to the formation epoch, and how much to subsequent
dynamical evolution.  There are now plausible theoretical and
observational grounds to suppose that the {\it formation} mechanism for
globular clusters in protogalaxies was largely independent of location in the
halo (\cite{hp94} [hereafter HP94]; \cite{mp96}; \cite{ee97}).  
But as soon as clusters form, they fall
prey to a variety of erosion mechanisms, especially tidal
shocking, early mass loss driven by high-mass stars, 
and (in disk galaxies) disk shocking.  In the innermost $\ltsim 2$ kpc,
dynamical friction may also play a role in removing the most massive
clusters (\eg, \cite{cpr96}; \cite{cap97a}).  All these mechanisms depend
sensitively on galactocentric distance $R_{gc}$ and the tidal field.  
Many theoretical discussions
(\cf\ the references cited above) have raised the possibility that the
inner-halo structure of the GCS might be completely dominated by dynamical
evolution, \ie\ that the great majority of the clusters originally present
there might now be gone.  For M87 particularly, several recent papers discuss
the possibility that the entire massive nucleus might have built up from
dissolved clusters (see \cite{cap93}; \cite{mcl95}; \cite{cap97}).  An
analysis specifically tuned to giant ellipticals such as M87 (\cite{mur97})
also suggests that both the total population and mass spectral index of the
GCS should suffer quite significant changes with time because the rates
of dynamical erosion depend on galactocentric distance.

On observational grounds, it is now clear that GCSs in large galaxies
display spatial distributions which 
typically have large {\it core radii}, typically a few kpc in size
(\eg, \cite{lau86}; \cite{hapv91}; \cite{gri94};
\cite{mcl95}; \cite{for96}).  It has frequently been
suggested that this feature might have resulted from
wholesale cluster erosion at small $R_{gc}$, a view which has been
argued by the dynamicists since the early paper of Tremaine
\etal~(1975).  Some recent continue to favor such
an interpretation (\eg, \cite{cap93} and \cite{cap97a}), while other
discussions favor considerably more modest effects 
of destruction (\eg\ \cite{mcl95}).
Clearly, the possibility that many clusters {\it might} have been 
destroyed in these inner regions does not require
that a large initial population of clusters {\it must}
have existed in the core regions of these galaxies; the question
is still open, and the direct
evidence that we are left with today is highly circumstantial at best.
An alternate approach (though one which has not yet been explored in
the same detail; see HP94) is that a similar result could have arisen
from the details of the cluster formation
epoch. For example, the strong tidal field could have inhibited the initial
buildup of globular-like protoclusters, 
or early mass loss could have evaporated
them rapidly in the first few $\sim 10^8$ y of their history.  Direct
observational evidence of cluster formation in the inner 
regions of the merger remnant NGC 7252 recently 
presented by Miller \etal~(1997) 
is intriguingly consistent with such a view, and deserves much 
closer evaluation as well as theoretical modelling.  

A somewhat more direct piece of evidence that dynamical evolution 
has dominated the construction of the GCLF that we now observe
-- in essence, something closer to a ``smoking
gun'' signature of erosive processes -- would be a progressive change in the
{\it shape} of the GCLF with galactocentric distance, \ie\ the relative
numbers of clusters of different mass.  If the initial mass distribution
of the clusters is at least roughly independent of galactocentric
distance (\eg, HP94; \cite{ee97}; 
\cite{mp96}; \cite{dif86}; \cite{ost97}), 
then such a change can only have arisen
from dynamical evolution.  On the theoretical side, this
is the approach taken by several other authors (\eg, \cite{cpr96}, 
\cite{ves97}, or \cite{ost97}).  We now need better and more extensive
data to compare with these simulations.

Within the Milky Way, the population of globular clusters at all $R_{gc}$ is
quite well defined over the full range of luminosities, 
but is too small a sample to provide strong statistical
tests (\eg, \cite{gne97}; \cite{kav97}).  M87 itself is perhaps the best
available testbed for these ideas, because of its simple gE structure,
relatively nearby location, and
enormous cluster population:  statistically useful numbers of globular
clusters can be traced in to within $\sim 1$ kpc of its center, and 
the GCLF can be studied to faint levels. 
In this paper, we present new photometric data for the
inner-halo GCS of M87 specifically designed to address these issues.

The paper is cast in two major parts.  In Secs.~2--4, we present
our new photometric data for M87, the luminosity function of the inner-halo 
GCS, and the color and metallicity distributions.  Then in Sec.~5, 
we take up the long-standing problem of understanding how M87 and other 
central-supergiant ellipticals can
generate such exceptionally large populations of globular clusters in all
parts of their halos (the ``specific frequency problem'').  
After assessing various suggestions in the literature, we conclude   
that the basic GCS characteristics were most likely built in at birth,
and suggest that they may have been influenced strongly by the early 
galactic wind in such galaxies.

\section{THE DATA}

\subsection{Observations and Calibration}

The observations for this study were obtained at the Canada-France-Hawaii
Telescope during 1993 March 28/29 and 29/30.  The High
Resolution Camera (\cite{mcc89}) was used at prime focus, with the
red-sensitive SAIC1 CCD detector:  this CCD (now decommissioned) had an 18
micron pixel size, a $1020 \times 1020$ format after bias subtraction and
trim, readnoise rms $5.5 e^-$, and gain 1.9 $e^-$ per adu.  With a scale of
$0\farcs 131$ per pixel in HRCam, the total field size was $2\farcm 23$ on a
side, corresponding to $9.7\times(D/15)$ kpc if M87 is at a distance of $D$
Mpc.  This size was comfortably enough to enclose the entire central region of
the M87 GCS, which has a core radius $r_c \simeq 1'$ (\cite{mcl95}).

For our deep GCLF observations, we used the $R-$band, which was the standard
filter nearest the peak sensitivity of the SAIC1 CCD.  The GCLF turnover (peak
frequency of cluster numbers per unit magnitude) is at $V \simeq 23.7$
(\cite{whi95}), or $R \simeq 23.2$ for $(V-R) \simeq 0.5$ (see below).  To
reach well past this limit while avoiding saturation from the bright
background light of M87, we used a long string of relatively short exposures
($26 \times 500$ seconds) with individual frames shifted along a sequence of 7
different positions separated by several arcseconds.  The individual exposures
were then re-registered and median-combined to construct a final deep image.
This image, which we will refer to as the ``central field,'' is shown in
Fig.~\ref{fig1} with the smooth galaxy light subtracted.  The field center
is actually almost $1'$ due south of the M87 nucleus, a choice which was
governed by the location of the best available nearby guide star to drive the
HRCam tip-tilt correction during the exposures.

Although the population of detected starlike images in the M87 central field
is totally dominated by globular clusters, an accurate correction for field
contamination requires a control field.  We selected a separate background
field located $15\farcm 09$ E and $0\farcm 45$ N of the central field and took
another series of deep $R$ exposures there.  At this projected distance from
the galaxy center, the number of M87 globular clusters is $\ltsim 2\%$ of the
level in the central field (see the wide-field GCS profiles of McLaughlin
\etal\ 1994).

Finally, we took shorter series of the central field in $V$ ($4 \times 600$
sec) and $I$ ($4 \times 500$ sec) to measure the colors of the brighter
globular clusters in the field.  These were used to estimate the size of the
GCS metallicity gradient in the core region.  A summary of the observational
material is given in Table 1.

Master flat fields in all three filters were constructed from a combination of
dome flats and twilight exposures, and the control field was used to check
that these produced globally flat calibrations to well within 1\%.  Absolute
calibration was accomplished with $VRI$ exposures of standard-star fields in
the clusters M67, M92, NGC 4147, and NGC 7006 (\cite{sch83}; \cite{chr85};
\cite{dav90}).  Both nights were photometric, and images of all
four standard fields were taken on each night.  The zeropoints of the
photometric scales in our M87 fields are judged to be uncertain by no more
than $\pm 0.02$ mag in each filter, from the observed scatter of the
bright-star aperture photometry used to transfer the calibrations.

\cite{cha90} obtained $BVI$ CCD photometry of $\sim 270$ clusters over a
roughly similar radial region around M87.  Unfortunately, we cannot compare
our data directly with theirs, since their field (just to the north of the M87
nucleus) has almost no overlap with ours.  The much wider-field study of
\cite{mhh94} in the $V$ band can, however, be used for comparison since
our outer zone overlaps with their inner one.  For 44 objects in
common brighter than $V \simeq 22.2$, we find $\Delta V$(MHH-HHM) = $-0.035
\pm 0.014$, confirming that there are no major zeropoint errors.

\subsection{Photometry}

To facilitate the photometry of the faint starlike objects in the central
field, we first removed the background light from M87 using the
ellipse-fitting codes in STSDAS.  Fig.~1 shows the field after subtraction of
the smooth and nearly circular isophotes.  As is apparent from the figure, the
model also successfully removed the bright semistellar nucleus of the galaxy,
leaving only the nuclear jet, the large population of globular clusters, and a
few faint background objects.

The photometry was done with the normal suite of DAOPHOT II and
ALLSTAR codes (\cite{ste92}) with a star-finding threshold set at $\simeq 3.5$
times the standard deviation of the average sky noise across the frame.  Two
iterations of the object-finding and ALLSTAR measurement were sufficient to
capture essentially all the starlike objects on the frames, since the field is
not crowded in any absolute sense.  Fig.~\ref{fig2} shows the locations of
the brighter ($R < 23.2$) objects on the frame, along with the field
orientation:  the $+x-$axis is $(32\fdg 95 \pm 0\fdg 5)$
eastward from due North.  A $\pm 15\fdg$ sector around the 
nuclear jet was excluded from any further analysis. 
A few clearly nonstellar objects (faint
background galaxies) were removed objectively from the detection lists by the
use of the image moment $r_{-2}$ defined in Harris \etal\ (1991).
Exactly the same rejection criteria were employed for the background field.
For the magnitude range $19.0 < R < 24.6$, 892 starlike objects in total
were measured in the central field and just 91 in the control field.
Table 2 contains a partial listing of the final data, for the brightest
objects:  here, $(X,Y)$ are in arcseconds relative to the center of M87, with
$X$ increasing eastward and $Y$ northward.  An electronic file of the complete
data may be obtained from WEH on request.

Since the pixel-to-pixel sky noise depends dramatically on radius from the
galaxy center, so does the completeness of detection at any given magnitude.
(Object crowding is not important at any radius, so the completeness $f$ is
determined almost purely by the background noise).  Extensive artificial-star
tests were carried out to determine $f$ and the internal precision of the
photometry:  scaled point-spread functions (psfs) were added to the frame,
typically in groups of $200-300$ over the entire magnitude range of interest,
and then subjected to exactly the same measurement procedure as the original
frame.  If $N(m)$ denotes the number of artificial stars inserted at magnitude
$m$ and $N_{meas}$ is the number successfully recovered (at any measured
magnitude), then we define $f(m) = N_{meas}/N(m)$.

After preliminary trials with various radial binnings, we divided the central
field into four annuli as listed in Table 3 and marked in Fig.~2.  The very
innermost circle ($r < 9'' \simeq 0.65$ kpc [D/15]) was eliminated from
further consideration because of its overwhelmingly high background noise,
which prevented any useful photometry.  The results for $f$, in steps of 0.5
mag, are shown in Fig.~\ref{fig3}.  Note that the control field reaches
significantly deeper than any part of the central field (despite its shorter
exposure time) because of its much lower background light.  In practice, we
approximate each of the curves in Fig.~\ref{fig3} by the two-parameter
Pritchet interpolation function (\cite{fle95})
\begin{equation} f(m) = {1 \over 2}\left[1-{{\alpha (m-m_0)} \over
{\sqrt{1+\alpha^2 (m-m_0)^2}}}\right]
\end{equation}
where $m_0$ (the ``limiting magnitude'' in a formal sense) is the magnitude at
which $f = 0.5$, and $\alpha$ measures the steepness of decline of $f(m)$ near
$m_0$.  The fitted values of $(\alpha, m_0)$ for each annular zone and the
control field are listed in columns (3) and (4) of Table 3.  Column (5) gives
the total area of each zone lying within the boundaries of the central field.

The artificial-star tests were also used to evaluate the internal photometric
precision, with the results shown in Fig.~\ref{fig4}.  Our limiting
magnitudes $m_0$ as listed in Table 3 correspond roughly to the levels at
which the measurement uncertainty reaches $\sim 0.25$ mag.

\section{THE LUMINOSITY DISTRIBUTION:  RADIAL TRENDS?}

McLaughlin (1995) has used our HRCam data to analyze the radial
structure of the inner GCS and to discuss its implications.  Here, we
investigate the characteristics of the GCLF (the luminosity function of the
globular clusters) in the GCS core and compare it with previously published
data for the outer halo.

Early studies of the GCLF in the Milky Way and M31, along with the first
photometry of the brighter clusters in the Virgo elliptical galaxies, gave
rise to the now-standard notion that the GCLF is roughly Gaussian in number of
clusters per unit magnitude (\eg, \cite{han77}; \cite{har79}; \cite{har91}).
On purely empirical grounds, the near-uniformity of the GCLF peak or
``turnover'' absolute magnitude has encouraged its use as a standard candle
(\cite{jac92}; \cite{har96b}; \cite{whi96}; \cite{kav97}).  The first
photometric study that penetrated unequivocally past the turnover point
in any galaxies beyond the Local Group was by Harris \etal\ (1991) for
the Virgo ellipticals NGC 4472 and 4649.  By now there are published studies for several
other large galaxies that reach clearly fainter than the GCS turnover level and
directly verify the near-Gaussian shape of the GCLF (\cite{fle95};
\cite{whi95}; Forbes 1996a,b).\footnote{It should be realized
that the Gaussian distribution is only a convenient fitting function
and has no astrophysical justification.  Other
equally simple analytic forms have been found that match at least as well
(\eg\ the $t_5$ function of \cite{sec92}). As Harris (1991)
points out, if the GCLF were truly Gaussian in form, then so too would be the
{\it luminosity-weighted} luminosity function (LWLF). However, the LWLFs 
are decidedly {\it non}-Gaussian (\cite{mcl94}; \cite{mp96}).}

In each of the four zones of our central field, the GCLF is defined by the number of
objects remaining after (a) removal of nonstellar images by image moment
analysis, (b) correction for detection incompleteness $f$, and (c) subtraction
of the background LF, also corrected for incompleteness and normalized to the
area of the zone.  The results are summarized in Table 4, in 0.4-mag bins
for each ring.  The first column gives the central $R$ magnitude of the bin,
and the next four columns list the residual number of objects in each zone,
fully corrected for incompleteness and background subtraction.  For
comparison, the last column gives the background LF (also corrected for
incompleteness).  From the relative zone areas listed in the last row, it can
be seen that the background corrections are much less than 10\% of the GCS
population in {\it every} zone and magnitude range except for the two faintest
bins in Ring 4.  Note also that we stopped the calculations in each zone when
the completeness $f$ dropped below $\simeq 0.4$, thus the effective
limiting magnitude increases with radius.

A potential concern is that our derived LFs might be contaminated
at the faint end by false detections from pure noise, particularly
in Ring 1 where the background light is high.  To measure this effect
explicitly, we inverted the image of the central field (\ie, we
reversed the sign
of all the pixel values on the sky-subtracted image) and then carried out
exactly the same photometric analysis on this inverted picture as on the
original one (see Harris \etal\ 1991 for a more extensive description of this
procedure).  The detected objects on this inverted frame are all 
pure noise by definition.  The vast majority 
are either fainter than our adopted limiting
magnitude in that radial zone, or are rejected as nonstellar in shape,
and the remaining corrections to the LFs are quite small, 
amounting on average to about 4 objects in the faintest bin in each zone.  
These corrections have been included in the final totals given in Table 4.

The GCLFs for the separate rings are plotted in Fig.~\ref{fig5}.  For rings
2, 3, and 4 the GCLF turnover (peak point) expected at $R\simeq 23.2$ has
clearly been reached and passed.  The question of immediate interest is
whether or not any systematic differences in the features of the GCLF show up
with galactocentric distance.  Standard Kolmogorov-Smirnov (K-S) two-sample
tests show that rings 2, 3, and 4 are {\it not} significantly
different from each other.  This result extends that of
\cite{mhh94}, who found no variation in the GCLF over the radial range
$1\farcm 2 \leq r \leq 6\farcm 8$.  The same conclusion can also be drawn from
attempts to fit Gaussian curves to the GCLFs, in which we find that the
turnover point $R_0$ and dispersion $\sigma(R)$ of the fitted Gaussians are
virtually identical for these three outer zones.  Combining rings $2-4$ for $R
< 23.8$, we find by weighted least squares that the best-fit Gaussian has $R_0
= 23.18 \pm 0.25$ and $\sigma = 1.39 \pm 0.15$.  This fit is displayed in
Fig.~\ref{fig6}.  Adding the mean color $(V-R) = 0.50$ (see below), we obtain
$V_0$(turnover) $ \simeq 23.7$, in complete agreement with the values $V_0 =
23.72 \pm 0.06$, $\sigma = 1.40 \pm 0.06$ quoted by Whitmore \etal\ (1995)
from an HST-based sample that reaches about half a magnitude deeper than ours.

For the innermost zone (ring 1), our data reach their reliable
limit just at $R \simeq 23.4$, and thus we cannot make any stringent test of
the GCLF turnover level except to conclude that $R_0$ is {\it no brighter}
there than in the outer zones.  Although Whitmore \etal\ (1995) do not comment
directly on this point, they also apparently did not find any strong 
changes in the turnover out to their radial
limit of $r(max) \simeq 114''$.  If we {\it assume} $R_0 \simeq 23.2$ for ring 1 as
well, then the data give a weak hint that the dispersion may be narrower
there, with a best-fit $\sigma \simeq 0.8 \pm 0.3$.  This result is indicated
schematically in the lower panel of Fig.~\ref{fig6}.  However, the relatively
small sample in ring 1 prevents us from strongly ruling out a higher $\sigma$.
A K-S test comparing ring 1 with either ring 4 or the sum of (2+3+4), over the
magnitude range $R < 23.4$, indicates that they are different at the $\simeq
90$\% confidence level -- again, suggestive but not definitive.

An alternate way of displaying the data to compare more readily with
theory is as $dN/dL$, the number of clusters per unit luminosity; we call this
function the LDF (HP94).\footnote{As noted by McLaughlin (1994) and
McLaughlin \& Pudritz (1996), the GCLF in its conventional form as number of
clusters per unit {\it magnitude} is just the luminosity-weighted first moment
of the LDF.} Theoretical formation models, as well as a large body of
observational material for currently active star- and cluster-forming regions
(\eg, HP94; \cite{mp96}; \cite{elm96}; \cite{ee97}), suggest that the initial mass
distribution function of the clusters should be closely approximated by a
power-law form $dN/dM \sim M^{-\gamma}$ where the mass spectral index $\gamma$
is typically in the range $\sim 1.5 - 2.0$.  The LDF will have the same
form as long as the mass-to-light ratio for individual clusters is independent of
$M$.  

We display the LDF in Fig.~\ref{fig7}, for three distinct
regions covering virtually the entire range of the M87 halo.  Our innermost
zone (ring 1) is shown as the upper set of points, 
covering the projected radial range
$\simeq 0.7 - 1.4$ kpc, while rings 2--4 ($\simeq 1.4 - 10.3$ kpc) are
plotted as the middle set of points.  Finally, for the LDF of the outer halo,
we use the $V-$band GCLF data of \cite{mhh94} covering the radial range
$2\farcm 88 - 6\farcm 82$ ($\simeq 12.5 - 29.8$ kpc) and shown as the lower
set of points.

The principal dynamical mechanism of tidal shocks should remove clusters more
efficiently at lower mass and thus progressively flatten the LDF over time --
and, of course, more rapidly at smaller $R_{gc}$.  How large is this
effect expected to be?   The Murali \&
Weinberg (1997) simulations starting from a very plausible initial mass
function indicate that $\gamma$ should decrease by $\sim 0.3$ over 10 Gyr of
evolution in an M87-like core.  In our data, we find that
over the luminosity interval $4.7 < {\rm log}(L/L_{\odot}) < 6.2$ (\ie, clusters
more luminous than the GCLF turnover point), direct weighted fits give $\gamma =
\Delta$log$N$/$\Delta$log$L = (1.59 \pm 0.16)$ for the
outer-halo zone, $(1.65 \pm 0.17)$ for the mid-halo, and $(1.84 \pm 0.48)$ for
the inner zone.  No significant differences emerge in any of the regions, and
if anything, the innermost zone exhibits a {\it steeper} mass function.  
This latter result is not an artifact of
increased background noise at faint magnitudes, 
since we explicitly measured the (small)
corrections due to noise and subtracted them out (see above).\footnote{The
larger uncertainty on $\gamma$ for the innermost zone is obviously a result of
its much smaller sample size.  Unfortunately, the sample cannot be increased,
since our data already include every cluster brighter than
the turnover point and within $r \ltsim 30''$ of the nucleus.  The outer-halo
sample of \cite{mhh94} also includes the great majority of the clusters
brighter than the turnover.  The only experimental prospect for drawing firmer
conclusions on the radial dependence of $\gamma$ 
is to obtain similar observations for many more {\it galaxies}.}
Our data, with their admitted limitations, therefore show no clear
evidence for the trends indicated by the models.

Alternately, we might ask more specifically about the effect of dynamical
evolution on the classic GCLF turnover point (plotted in its normal
observational form as number per unit magnitude).  The progressive
removal of low-mass clusters in the inner regions, as noted above,
should cause the turnover point to become brighter with time
and the GCLF dispersion to become narrower.  The most recent numerical
simulations which deal with the GCLF 
(Capriotti \& Hawley 1996, Vesperini 1997, Ostriker \& Gnedin 1997)
find that the GCLF {\it shape} tends to be roughly preserved 
as evolution continues, once it approaches its present Gaussian-like form,
and that the evolution in the turnover point can be surprisingly modest.
In particular, Ostriker \& Gnedin 
(1997) predict that the cluster populations in 
the inner $\sim 5$ kpc of large galaxies like M31, M87,
and the Milky Way will have turnover points that are brighter by
typically ($\sim 0.3 \pm 0.1$) magnitude than the outer-halo
populations in the same galaxies, along with narrower GCLF dispersions
by typically $0.1 -- 0.2$ magnitude.  These radial differences -- both
rather modest -- are at least weakly consistent with our observations,
which show similar turnover levels to within $\pm0.2$ in all of our
zones, but a narrower dispersion in the innermost zone.
We remark that Kavelaars \& Hanes (1997) find a rather similar
observational result for the Milky Way and M31:  the inner-halo GCLFs
are more sharply peaked, yet have turnover luminosities that are
statistically indistinguishable from those in the outer halo.  An earlier
version of this same conclusion was found by \cite{arm89}, and the steady
improvements in the Milky Way database have reinforced it.

In summary, (i) the modern theoretical simulations predict rather modest
effects on the GCLF with radius (at least for the brighter part of the
GCLF that is most easily observed), and (ii) our data indicate
trends with radius that are, if anything, even smaller than the model 
predictions.
The destructive mechanisms described above are important primarily
on clusters {\it fainter} than the nominal turnover (log $L/L_{\odot}
< 4.7$, or $M \ltsim 10^5 M_{\odot}$).  Thus to find more unequivocal
evidence for dynamical effects, we will need
to explore the GCLF to much fainter magnitude levels and at widely separated
places throughout the M87 halo.  Note, however, that the deepest available
study (\cite{whi95}) shows no clear evidence for radial variations in the GCLF
for $M_V \ltsim -5.4$, corresponding to $M \gtsim 3 \times 10^4 M_{\odot}$.

A more extreme -- and more contrived -- possibility is that
we are witnessing the effects of convergent evolution, in the
sense that the {\it initial} cluster mass distribution was steeper in the
core, and has just now reached a flatter slope which coincides with the more
slowly evolving outer-halo LDF.  
Such an explanation seems unlikely to us particularly because it would
have to apply equally well for many different 
types of galaxies (dwarfs, spirals,
and ellipticals) within which the detailed destruction rates would
differ.  Also, the formation models of McLaughlin \&
Pudritz (1996) and Elmegreen \& Efremov (1997), 
which successfully account for the power-law shape of the LDF, 
already give strong grounds for expecting
that the initial LDF should be independent of galactocentric radius.

Another notable feature of Fig.~\ref{fig7} is the very top end of the LDF (log
$L/L_{\odot} > 6$, equivalent to $M \gtsim 2 \times 10^6 M_{\odot}$), which   
cuts off more sharply for the middle and inner zones than for the
outer halo; that is, the {\it very} most massive globular clusters seem
progressively to disappear as we move inward.  Is this feature showing us
evidence of dynamical friction, which is expected to become important for
exactly this mass range (\cite{tre75}; \cite{cap97a})?  
The recent simulations of Capriotti
\& Hawley (1996) for clusters in a large-galaxy potential 
predict that most of these massive clusters ($10^6 - 10^7
M_{\odot}$) should survive if $R_{gc} \gtsim 20$ kpc; about
half will be destroyed near $R_{gc} \sim 8$ kpc; and almost all will disappear
within $\simeq 2$ kpc.  For M87, we find from Fig.~\ref{fig7} that
there are $19\pm8$ clusters 
brighter than $V \simeq 20.25$ ($\log\,L/L_{\odot} \simeq 6$) in the
outer zone; $4\pm3$ clusters in the
middle zone; and none in the inner zone.  However, these numbers must be
normalized to the same parent population:  the total number of clusters over
the range $20.25 < V < 23.25$ is $\simeq 650$ in the
outer zone, 350 in the middle zone, and 40 in the inner zone. Thus in
proportion to these totals, we would have expected to see $10\pm4$ of the
super-bright clusters in the middle zone if none were destroyed, but about
half this many if the models are correct;  we observe $\sim 4$ in the
real sample.  In the inner zone, we would expect just one cluster if no
dynamical destruction has occurred; we see none.  Although these comparisons
are only approximate ones (the radial binnings in the models are rough ones,
the statistics are weak, and the observed numbers are for projected radial 
bins rather than true three-dimensional radii),
the model predictions are consistent with the data.

Another way to state this conclusion is that, for the radial
regime ($r \gtsim 1$ kpc) that we can most easily observe, dynamical
friction has had little effect on the cluster LDF.  We therefore concur
with Lauer \& Kormendy (1986) and McLaughlin (1995) that the GCS core
radius, which is $\gtsim 4$ kpc, is too large to have been created 
by the action of dynamical friction on a hypothetical 
GCS that initially followed
the more centrally concentrated stellar light profile ({\it unless}
the inner clusters all formed in severely radial, plunging orbits that
took them well within the stellar bulge; see \cite{cap97a}, \cite{cap97}.
However, we regard this as a highly speculative alternative lacking any direct
evidence; see the papers cited for additional discussion).

Interestingly, to detect the effects of dynamical friction on the
lower-mass parts of the LDF more directly (\ie, $M \ltsim 10^6 M_{\odot}$), 
the models cited above indicate that we would
have to study the cluster mass distribution within
a radial regime that penetrates into the core radius
of the M87 stellar {\it bulge} light, \ie\ to within $r \ltsim 0.5$ kpc
(or about $7''$).  Unfortunately, there appear to be 
no immediate prospects for
doing this:  at such small radii, it is currently impossible to find and
measure the fainter clusters, and there are virtually no clusters present
in the first place within this small projected area.
In other galaxies, the GCS populations are much
smaller than in M87, and the situation is therefore even 
worse on observational grounds.

In conclusion, we favor the view that dynamical evolution has {\it not} 
had very dramatic effects on the shape of the GCLF for the mass range $M
\gtsim 10^5 M_{\odot}$, except possibly 
at the very top end ($M \gtsim 2 \times 10^6
M_{\odot}$) where dynamical friction may have 
cut off the most massive clusters
within a few kpc of the nucleus.  The best present observations are
beginning to converge with current theoretical simulations.  
Still deeper measurement of the cluster population at
several different places in the M87 halo would be extremely valuable,
with the potential to place still more stringent limits on the
dynamical models.  

\section{COLOR AND METALLICITY DISTRIBUTIONS}

The shorter exposures of the central field in $V,I$ (Table 1) were used to
measure color indices for the brighter M87 clusters and thus to investigate their
metallicity distribution.  The color-magnitude diagrams from our data are
shown in Fig.~\ref{fig8}.  From the luminosity function of the control field
(Table 4), we estimate that just 10 objects brighter than $R \simeq 23$ should
be due to field contamination, so the points plotted in these diagrams are
almost entirely M87 globular clusters. The overall mean color indices of the
sample (for 155 clusters brighter than $R = 22$) are $\langle V-R \rangle =
0.501 \pm 0.008$ with rms scatter $\sigma = 0.10$ mag; $\langle R-I \rangle =
0.617 \pm 0.011$ with $\sigma = 0.13$ mag; and $\langle V-I \rangle = 1.118
\pm 0.008$ with $\sigma = 0.17$ mag.  The scatter due to photometric
measurement uncertainties alone averages $\pm 0.07$ mag in each filter over
this magnitude range. The intrinsic dispersion in cluster color should then
be $\sim 0.15$ mag in $(V-I)$, the index most sensitive of the three to
metallicity.

Recently, Elson \& Santiago (1996a,b) and Whitmore \etal\ (1995) have obtained
HST $(V,I)$ photometry for samples of M87 globular clusters from three
different regions in the halo.  They find a distinctly bimodal color
distribution -- previously suspected to exist by \cite{lee93} from $(C-T_1)$
photometry with somewhat lower internal precision -- with histogram peaks at
$\langle V-I\rangle \simeq 0.92$ and 1.23.  These colors correspond (see
below) to [Fe/H] $\simeq -1.7$ and 0.0. In addition, at
high metallicities, all of the conventional photometric indices may give
systematically incorrect metallicity values (additional comments will be made
on this point below).
Regardless of the exact metallicities of each of the two groups, bimodal
distributions of this type have the traditional interpretation
that the halo formed in two major enrichment stages, 
involving comparable amounts of gas.  Very similar bimodal
distributions are clearly present in the GCSs of many giant ellipticals (\eg,
\cite{for97}; \cite{gei96}; \cite{zep95}) as well as in the Milky Way (Zinn
1985; Armandroff 1989), but may be absent, or considerably less obvious,
in several other ellipticals (\cite{ajh94}; \cite{for96}).  Bimodality thus 
seems to be a common, but perhaps not universal, phenomenon (but see
\cite{cot97}).

Elson \& Santiago (1996a,b) also claim the existence of a mild trend in mean
metallicity with magnitude in the sense that the bluer (more metal-poor)
clusters tend to be brighter.  Similar trends are not obviously present in the
much larger samples of clusters measured by Whitmore \etal\ (1995) or
\cite{lee93}, and Forbes \etal\ (1997) do not find any such
effect in another giant elliptical, NGC 5846.  Careful inspection of Elson \&
Santiago's data (\cf\ Fig.~2 from their second paper) shows that their
claimed trend in fact arises from an excess population of clusters in the
``blue'' group ($V-I \simeq 0.92$) concentrated near $V \simeq 22.0$.  This
extra little clump of objects is {\it not} near the top end of the GCLF,
despite its appearance in the Elson-Santiago diagrams; it is actually almost
two magnitudes fainter than the GCLF tip.

Our sample of clusters is drawn from a parent population more than three times
larger than the Elson-Santiago sample, so we should be able to make a stronger
test for the reality of this feature.  In Fig.~\ref{fig9}, the mean $(V-I)$
color is plotted as a function of magnitude in $0.5-$mag bins, extending to
the faint limit of our data.  Our conclusions from this exercise are that (a)
the {\it very} brightest clusters (based on just 7 objects brighter than $R =
20$) tend to fall in the ``red'' (more metal-rich) group; and (b) for $R > 20$,
there is no statistically significant change in mean color with magnitude.
Nevertheless, we detect the Elson-Santiago feature as a slightly
bluer than average color (by 0.05 mag) at just the magnitude level that
they noticed ($R \simeq 21.5$, $V \simeq 22.0$).  In conclusion, we verify the
existence of this mildly anomalous feature, but we do not find that it affects
the overall structure of the GCLF in any important way.  Nevertheless, their
study raises the possibility of interesting fine structure in the
color-magnitude diagram which should be pursued with 
higher precision photometry of a larger sample.

Finally, we use our $(V-I)$ measurements to investigate the overall {\it
metallicity gradient} in the M87 halo.  Our material, covering the inner halo,
can be combined with that of \cite{lee93} for the outer halo to give us a
comprehensive view of the entire galaxy.  To convert $(V-I)$ into metallicity,
we assume $E(V-I) = 0.03$ (\cite{bur84}) for M87 and use the conversion relation
\begin{equation}
(V-I)_0 = 0.18\,{\rm [Fe/H]} + 1.20
\end{equation}
which we derive from the recent database of colors and metallicities for the
Milky Way clusters (\cite{har96}); it is nearly identical with earlier
calibrations (Couture \etal\ 1990; 
\cite{kis97}), and is based essentially on the \cite{zin85}
metallicity scale for the Milky Way clusters. We note in passing that these
photometrically derived metallicities depend critically on the assumption that
this simple linear relation is valid even at metallicities [Fe/H]
$\sim 0.0$ and higher, even though there are no Milky Way globulars with
accurately measured metallicities this high.  
This simple assumption may not be correct and that the extrapolation of this
relation may seriously overestimate the true metallicities for [Fe/H] $\gtsim
-0.5$ (see \cite{har92}; \cite{car97}).  

Our results for the mean metallicities are listed in Table
5, for seven radial bins with $\simeq 20$ clusters per bin:  column (1) gives
the mean radius of the clusters in the bin, column (2) the number of objects
falling within the broad range $0.8 < (V-I) < 1.5$, columns (3) and (4) the
mean color and the rms scatter of the sample, and columns (5) and (6) the
mean metallicity and dispersion.  These points are plotted in
Fig.~\ref{fig10}, in which the error bars shown are the standard
deviations of the mean metallicities directly from Table 5.  The
{\it zero point} uncertainty of the [Fe/H] scale (\ie, the external error)
is expected to be $\ltsim \pm 0.25$ dex, from the $\pm 0.03-$mag
uncertainty in the photometric zeropoint of our $(V-I)$ scale (\S 2.1
above) and the similar uncertainty in the coefficients of the conversion
relation Eq.~(2).  In Fig.~\ref{fig10},
we also plot the \cite{lee93} metallicity
measurements derived from their $(C-T_1)$ index.  Encouragingly, in the radial
range of overlap between our data and theirs ($r = 60'' - 100''$), the mean
metallicity values agree to well within $\pm 0.2$ dex in [Fe/H],
suggesting that the calibrations of both photometric metallicity scales
are self-consistent to the level we expected.

The total range in radius covered by Fig.~\ref{fig10} now extends from
$10''$ to $500''$ (or 0.7 kpc to 36 kpc).  It is apparent that the
{\it mean} GCS metallicity decreases steadily outward from $r \simeq 60''$ to
the outermost limits of the data, even though the {\it range} of cluster
metallicities is comparably broad at any given radius (Lee \& Geisler 1993;
\cite{whi95}; Elson \& Santiago 1996a,b).  Although the scatter of the mean
points does not permit any very precise determination of the slope, the rough
relation
\begin{equation}
{\rm [Fe/H]} \simeq -0.9\ {\rm log}\,r'' + 1.15 ~~~ (r > 60'')
\end{equation}
adequately describes its systematic change. An interpretation suggested by
Geisler \etal\ (1996) from their extensive analysis of the NGC 4472 GCS, which
is quite similar to that of M87, is that the bimodal metallicity distribution
is present at all radii and that there is little, if any, metallicity gradient
{\it within} either of the two subsystems. However, the metal-richer ones form
a much more centrally concentrated subsystem (as they do in the Milky Way), so
their relative numbers increase strongly inward and thus create the observed
shift in mean metallicity with radius.  Notably, however, we find that {\it
within the GCS core radius} of $r_c = 1' \simeq 4.3$ kpc, the
mean metallicity stays constant with radius.

\section{THE FORMATION OF SUPERGIANT ELLIPTICALS}

\subsection{M87 and the Specific Frequency Problem}

The preceding discussion indicates that for M87, the inner-halo part of the GCS 
is simply an inward extension of the rest of the halo:
the GCS in the core has a luminosity distribution function scarcely different
from that of the outer halo, and the core metallicity distribution has a mean
[Fe/H] $\sim -0.3$ which would result from its being drawn primarily from
the metal-richer part of the bimodal distribution, along with a smaller
admixture from the metal-poor component.
All the available data are quite consistent with the picture 
that the internal characteristics of the M87 GCS 
exhibit no abrupt changes with radius.

To this point, we have not yet brought in another important
characteristic of the M87 system, which is the {\it specific frequency}
$S_N$ (\cite{hvdb81}; \cite{har91}), or number of clusters per unit
galaxy luminosity.  The specific frequency is
\begin{equation}
S_N = N_{GC} \cdot 10^{0.4(M_V^T + 15)}
\label{eq4}
\end{equation}
where $M_V^T$ is the integrated magnitude of the host galaxy and $N_{GC}$
is the total number of globular clusters in the galaxy.
It has long been realized that $S_N$ correlates
strongly with environment:  the highest-$S_N$ systems
are found only in some cD-type giants at the centers of rich galaxy
clusters (\cite{har95}; \cite{bla97}\ [hereafter B97]), while 
the sparsest globular cluster populations usually reside in ``field'' 
or small-group ellipticals in lower-density environments (\cite{har91}; 
\cite{wes93}).  Thus to develop a larger picture of the 
origin of the GCS, we must also connect M87 to similar central-giant
ellipticals in other clusters.  We now turn to this somewhat broader issue.

We first provide an improved calibration of $S_N$ for M87, by combining
our new data with the wider-field counts of McLaughlin \etal\ (1993,
1994) and Harris (1986).   McLaughlin \etal\ find 
$3100\pm130$ clusters brighter than $V=24$ over  projected radii $1\farcm 97\leq
r \leq 9\farcm 09$, while the GCS radial profile of McLaughlin (1995)
from our HRCam data 
gives $1300\pm80$ clusters brighter than $V= 24$ and within $r = 1\farcm 97$.
For the GCLF parameters adopted above ($\sigma = 1.4$, $V_0 = 23.7$),
the globulars brighter than $V=24$ account for $58\% \pm
6\%$ of the whole population; thus, the total number of M87 clusters with
$r\leq 9\farcm 09=39.7(D/15)$ kpc is $7590\pm1050$.  Then from the M87
surface photometry of \cite{dvn78}, we find that the integrated magnitude of
the galaxy interior to $r=9\farcm 09$ is $M_V=-22.3$ [for $(B-V)=1.0$
and $(m-M)_V=31.0$]. The {\it metric specific frequency} of M87, which is
defined by B97 as the ratio of cluster numbers to galaxy light inside $r=40$ kpc,
is then
\begin{equation}
S_N^{40}(M87) \simeq (7590\pm1050)\times 10^{0.4(-22.3+15)}=9.1\pm1.3\ .
\label{eq5}
\end{equation}

The number density of globular clusters in M87 falls off more gradually with
$r$ than does the intensity of background galaxian light (\cite{hs76};
\cite{har86}; \cite{mhh93}), so the specific frequency must increase with
galactocentric radius and the global $S_N$ of the
entire galaxy necessarily exceeds $S_N^{40}$. To estimate the
former quantity, we adopt $r=25\arcmin = 109(D/15)$ kpc as the 
``edge'' of the galaxy (which has a total magnitude of $M_V^T =-22.43$; de
Vaucouleurs \& Nieto 1978) and of the GCS. The observed surface density
profile of the outer cluster system 
can be taken from the wide-field counts
of Harris (1986); these show that the cluster population beyond $9'$
makes up $(1.048 \pm 0.081)$ times the sum from $r=2'$ to $9'$.  The total number of
clusters over all radii and magnitudes is then $13200 \pm 1500$, and hence
\begin{equation}
S_N({\hbox{M87}})=14.1\pm1.6\ .
\label{eq6}
\end{equation}
The main source of uncertainty is the behavior of $\sigma_{\rm cl}$ in the
outermost halo of M87.

The global specific frequencies of other elliptical galaxies in
Virgo are typically of order $S_N \sim 5$ (\cite{har91}).  A valuable 
template for a `normal' galaxy in this respect is M49 = NGC 4472,
which is comparable to M87 in brightness and size, but which has a
global $S_N=6.0\pm1.5$ and a metric $S_N^{40}\simeq4\pm1$ (from the data of
\cite{har86}).  Over all types of environments, $S_N$ for E galaxies
ranges from a minimum below
$S_N \sim 1$ in some field ellipticals up to $\gtsim 15$ for the richest cD's.
The underlying cause for this amazingly wide range among otherwise
similar galaxies has been rather a mystery since the phenomenon first became
clear two decades ago (\eg, \cite{vdb77}; \cite{har78}; \cite{hvdb81}).

\subsection{Scaling Relations for Brightest Cluster Galaxies}

The search for an empirical set of necessary {\it and sufficient} 
conditions for high $S_N$ has been a long one. 
While the highest$-S_N$ galaxies often have a cD morphology, and
are always the central members of their parent clusters or subclusters,
there are many other cD and brightest cluster galaxies (BCGs) with quite
normal GCS populations (\eg, McLaughlin \etal\ 1994; Harris \etal\ 1995). 
However, the recent work of B97 and \cite{btm97} (hereafter BTM97) considerably
enlarges the database of GCS properties for these galaxies.
Blakeslee \etal\ show that $S_N^{40}$ in the BCGs
is correlated with several indicators of the total cluster mass, such as the
velocity dispersion of the galaxies and the X-ray luminosity of the hot gas
residing in the cluster potential well.  Moreover, it is now apparent 
from the BTM97 study that {\it BCGs cover a smooth continuum} of specific 
frequency values, from normal levels up to the most extreme known.

First we collect the available data for these BCGs.
BTM97 give the ``metric'' $S_N^{40}$ values for BCGs 
in 19 Abell clusters, measured by their surface brightness
fluctuation technique.  In addition, Harris \etal\ (1995) give global $S_N$ values 
for other BCGs measured by direct resolution 
of the GCS population.  From seven galaxies
measured through both techniques (NGC 1399, 3842, 4486, 4874, 4889, 6166,
and 7768), we derive a mean ratio $\langle S_N/S_N^{40}\rangle = 
1.3 \pm 0.2$ which can be used to convert approximately between these two 
quantities in cases where only one has been measured.
In Table 6, we summarize the updated parameters for 11 centrally dominant
giant E galaxies in clusters not listed by BTM97.
Here, columns (1) and (2) give the galaxy name and host
cluster; column (3) the total luminosity of the BCG (assuming
$H_0 = 80$ km s$^{-1}$ Mpc$^{-1}$); columns (4) and (5)  the
global and metric specific frequencies; column (6) the velocity dispersion
$\sigma_{cl}$ of the galaxies in the cluster; and column (7) the X-ray 
temperature of the hot ICM gas in the cluster.  Reference
sources are listed in parentheses in columns (6-7).  Adding them to the 
Blakeslee sample, we now have a total of 30 BCGs 
spanning the entire range from quite
sparse groups (Leo, Centaurus) all the way up to the richest types of
Abell clusters.\footnote{In a few cases, the BCG
is actually not the optically brightest galaxy in the cluster (Virgo and
Coma are well known examples in which another cluster elliptical is
slightly more luminous than the central one).  Here, we are using the 
term ``BCG'' specifically to mean the {\it centrally located} 
giant E galaxy in the cluster.}

The basic difference between the BCGs and other E galaxies is shown
graphically in Fig.~11.  Here, the total population of the GCS
(calculated from the global $S_N$ and the galaxy luminosity)
is plotted against $M_V^T$.  A total of 82 elliptical galaxies are 
now included in this plot; data for the non-BCG systems are taken from 
the compilations of \cite{hh97} and Durrell \etal\ (1996) 
with a few more recent updates from the literature.  
Despite the significant scatter, it is clear that 
to first order, $N_{GC}$ is directly proportional
to galaxy luminosity (that is, $S_N \simeq$ constant) over a range of more
than $\sim 10^4$ in $L_{gal}$.
However, {\it for the BCGs alone, $S_N$ is itself a function of galaxy size}, 
increasing steadily toward larger and more massive systems.

As is clearly shown by B97 and BTM97,
BCG specific frequency also increases with 
total cluster mass (represented either
by the velocity dispersion $\sigma_{cl}$ or the ICM gas temperature $T_X$).
In Fig.~\ref{fig12} we show $S_N$ versus $T_X$ for our augmented sample
of BCGs, and Fig.~\ref{fig13} is a similar plot of $S_N$ and $N_{GC}$
versus $\sigma_{cl}$.  The overall dispersion  of points in
both graphs shown here -- as well as the several others shown in BTM97 -- 
is larger than for the Blakeslee sample by itself.  Nevertheless,
the general trend remains that more populous
GCSs {\it and} higher specific frequencies are found 
in galaxy clusters with systematically higher mass, deeper potential wells,
and hotter and more massive ICMs.
Probably the clearest representation of these trends appears in the plots
against $\sigma_{cl}$.  As did BTM97, we adopt $\sigma_{cl}$
as the primary indicator of the size of the cluster, and using this, we can
derive several simple
{\it empirical scaling relations} for specific frequency, total GCS
population, and BCG luminosity.  Least-squares solutions
against cluster velocity dispersion give the following:
\begin{equation}
{\rm log}\, S_N = (0.71 \pm 0.26)\, {\rm log}\, \sigma_{cl} - (1.11 \pm 0.68) \, ,
\label{eq16}
\end{equation}
\begin{equation}
{\rm log}\, N_{GC} = (1.58 \pm 0.32)\, {\rm log}\, \sigma_{cl} - (0.42 \pm 0.86) \, ,
\label{eq17}
\end{equation}
\begin{equation}
{\rm log}\, (L_{gal}/L_{\odot}) = (0.86 \pm 0.19)\, {\rm log}\, \sigma_{cl} + (8.55 \pm 0.51) \, ,
\label{eq18}
\end{equation}
where $\sigma_{cl}$ is in km s$^{-1}$.  In other words, we have approximate
scaling laws $N_{GC} \sim \sigma^{1.6}$,
$L \sim \sigma^{0.9}$, and $S_N \sim N/L \sim \sigma^{0.7}$.
(We ascribe no particular significance to the fact that these correlations
all pass through $(0, 0)$; at the low-mass end, all these relations require
downward extrapolations which are not constrained by our data.)
These relations stand independently of any of the theoretical hypotheses
that will be discussed below.

Much of the scatter in all these correlations may simply be due to the 
sizable measurement uncertainties in the individual $S_N$ values. 
Only two objects stand out as genuinely anomalous:
these are NGC 1399 (a high$-S_N$ but low-luminosity 
cD in the relatively low-mass Fornax cluster), and NGC 1275 (the extremely luminous
but {\it low}$-S_N$ central cD in the massive Perseus cluster).  
The populous GCS in NGC 1399 has been measured
several times with highly consistent results 
(\cite{hh86}; \cite{bri91}; \cite{kis97}, among others),
and the galaxy's luminosity profile is well enough established 
(\eg\ \cite{kil88}) to permit accurate
calibration of either $S_N^{40}$ or $S_N$(global).
NGC 1399 is a challenging anomaly in almost any formation scenario.
In the case of NGC 1275, the central galaxy 
is clearly extremely luminous even after accounting 
for the recent star formation present there, and the GCS is
just as clearly unexceptional (\cite{kai96}).  NGC 1275 might 
plausibly be explained by supposing that the Perseus cluster 
is a rare instance where the central cD built up from an initially 
rather normal E galaxy through amalgamation with many neighboring
galaxies, as discussed by Kaisler \etal\  

\subsection{Evolutionary Scenarios}

How did these intriguing BCG systems arise?
The ideas that have developed in the literature can be grouped into
three main approaches: 
(1) the high-$S_N$ galaxies acquired large
numbers of ``extra'' globular clusters in various possible ways long after their
initial formation stage; or
(2) the high $S_N$ is a form of optical
illusion, in which a population of intergalactic globular clusters that simply
occupy the potential well of the cluster as a whole lies at
the same location as the BCG; or 
(3) the GCS populations were largely built in at birth, with later minor
adjustments.   In the next three sections, we discuss 
each of these options in turn.  A key to keep in mind
is that any proposed solution to the $S_N$ problem must explain not only 
the sheer total number of clusters present in these galaxies; it must also
address several other measurable GCS characteristics including their
spatial distribution, velocity distribution function, and metallicity 
distribution function.  As yet, no complete model has been developed
which can quantitatively deal with all these aspects of the problem.
To help evaluate the various options, we first comment on the evolutionary
scenarios.

Among the proposed mechanisms for supplementing the GCS populations of
dominant cluster galaxies relatively late in their lives are the following:

\noindent (1) Cooling flows:  M87 and other cD galaxies have extended coronae
holding large amounts of hot low-density gas, and it has
been hypothesized that globular clusters somehow actively form out of the
ongoing cooling flow from this gas, thus adding to the original cluster
population (\cite{fab84}).  However, this model fails on several grounds to
predict the quantitative features of the observed GCSs in these galaxies,
such as the cluster metallicity distribution; the complete lack of
correlation between GCS population and cooling-flow amplitude;
the near-complete absence of any young, massive star clusters in most
cooling-flow galaxies; and the large velocity dispersion in the GCS
which is unlike that expected from objects condensing out of a cooling flow
(see \cite{gri94b}; Harris \etal\ 1995; \cite{kai96}; \cite{bri96}; \cite{hol96}).

\noindent (2) Stripping from other galaxies:  the possibility that central
BCGs might have acquired halo globulars from other, {\it fully formed}
galaxies during close encounters was explored in a series of models by Muzzio
and collaborators (see \cite{muz87} for a review).  These investigations showed
that cluster swapping may be responsible for minor rearrangements of clusters
between galaxies, but is unable to produce the major increases in $S_N$
(factors of 2 to 3) that we need for the central cD galaxies. A related
possibility, in which tidal debris from systems throughout a galaxy cluster
sinks onto a BCG at the bottom of the potential well (again, after the main
epoch of galaxy formation), is discussed and rejected by \cite{mhh94}.

Another version of this approach, pointed specifically at explaining
the bimodal metallicity distributions seen in giant elliptical GCSs,
has recently been developed by C\^ot\'e etal\ (1998).  They 
assume that a typical
giant elliptical such as NGC 4472 started as a `seed galaxy' of smaller
size, and has since absorbed large numbers of smaller (mostly dwarf)
galaxies.  The metal-richer globular clusters in the final gE are thus
mostly ones that formed with it originally, while the metal-poorer
clusters are the ones acquired from the accreted dwarfs.  
By quantitative numerical simulations, C\^ot\'e \etal\ 
show that a plausible range of GCS metallicity distributions
can be accounted for this way.  To account for a much higher-$S_N$ BCG
like M87, they assume further that the giant central galaxy accretes only
the high-$S_N$ envelopes of the smaller galaxies orbiting in the
potential well of the galaxy cluster, without absorbing the central
(cluster-poor) cores of those galaxies.  The advantage of this general 
scheme is that it employs a process (small-galaxy 
accretion and harassment) which must 
surely happen at some level in almost any rich environment like
Virgo, Fornax, or Coma.  However, some related worries are that
(i) to explain the {\it entire} GCS metallicity distribution function
this way, it is necessary to assume that very large numbers of small
galaxies are accreted (many hundreds for a case like NGC 4472, and even
more for M87); (ii) if (typically) half the light of the gE is from
smaller, accreted, and more metal-poor systems, they should not have
the high ($\sim$ solar or higher) mean metallicities that they are
observed to have; and (iii) an environment like Virgo should contain
hundreds (perhaps thousands) of stripped cores of dwarf galaxies whose
envelopes were accreted by the central BCG.  These expectations run
counter to the available observations.
Nevertheless, their discussion correctly calls for additional
investigation through detailed N-body simulations.

\noindent (3) Mergers:  E galaxies can be built out of the collisions and
mergers of disk galaxies, and massive globular-like star clusters can form in
the shocked gas during the merger (Schweizer 1987, 1997; \cite{kum93a}).  The newly formed
clusters may end up in a more centrally concentrated and more
metal-rich subsystem than the original (metal-poor, spatially extended) GCSs
of the colliding galaxies (\cite{ash92}; \cite{zep93}). Several galaxies in
which this process is actively happening have now been identified (\eg,
\cite{hol92}; \cite{whi93b}; \cite{ws95}; \cite{hol96}; \cite{mil97}).  
It is important to keep in mind
that the specific frequency of the merged remnant may end up {\it either}
larger or smaller than in the pre-merger galaxies, depending on factors such
as the total amount of incoming gas, the collision geometry, and (most
importantly) the efficiency of cluster formation relative to field-star
formation.

The merger mechanism is arguably quite a plausible way to build ellipticals
that end up with low-to-normal $S_N$ (Harris 1981, 1995; \cite{mil97}), 
since typical disk
galaxies have relatively few halo clusters of their own, as well as modest
supplies of gas.  For example, isolated E Galaxies such as NGC 3557
(\cite{mor85}) and NGC 720 (\cite{kis96}), which have elongated shapes and
very low specific frequencies in the range $S_N \ltsim 2$, are prime
candidates to be the results of disk-galaxy mergers that happened long ago.
The more recent merger remnants such as NGC 3921 (\cite{sch96}) and
NGC 7252 (\cite{mil97}) appear to be 
evolving into the same type of elliptical, 
with modest populations of globular clusters and $S_N \sim 2-3$,  
as predicted by Harris (1995).
However, applying the same model to giant ellipticals at
the highest end of the $S_N$ scale leads to much more serious difficulties.
For such galaxies, thousands of massive star clusters would have to be
formed during the mergers, requiring an enormous input supply of gas -- so
large, in fact, that the merging ``galaxies'' must in essence be pregalactic
clouds (\cite{har95a}).  In addition, some high-$S_N$ galaxies do not have
clearly bimodal GCS metallicity distributions; the highest-$S_N$ systems do
not have the lowest radial metallicity gradients; and the lower-metallicity
clusters in these gE's are often not as metal-poor as those in spirals.
All of these points -- and others raised in
the references cited above -- speak against the merger
model as a viable origin for the extreme BCGs (\cite{gei96}; \cite{for97}).  

In summary, these secondary-evolution processes appear able 
to produce measurable
changes in $S_N$ for ellipticals with cluster populations in the low or normal
range.  However, for M87 and the other supergiant cD galaxies with the highest
specific frequencies, no viable production mechanisms have 
convincingly emerged from these approaches.  

\subsection{Intergalactic Globular Clusters}

\cite{whi87} suggested that clusters of galaxies might host populations of
globular clusters that are not currently bound to any one galaxy, but instead
move freely in the potential well of the cluster as a whole. \cite{wes95} have
revived this idea, and cite what is essentially a version of
Blakeslee's (1997) correlation between BCG specific frequency and galaxy
cluster X-ray luminosity as possible evidence for the existence of these
intergalactic globular clusters (IGCs). In their scenario, the $S_N$ of a
central BCG could be  artificially enhanced by the superposition
of IGCs --- which would naturally have their largest space density in the very core
of the cluster --- on top of a normal ($S_N\sim 5$) population of globulars
that actually are bound to the BCG. 
Indeed, given the existence of diffuse light in galaxy
clusters (\eg\, Coma: \cite{thu77}), 
and the recent identification of candidate
intergalactic stars in Fornax (\cite{the97}) and Virgo (\cite{arn96};
\cite{fer96}; \cite{men97}; \cite{cia98}), it would be surprising 
if IGCs do {\it not} exist. However, the
question whether a population of IGCs could {\it single-handedly} give rise
to the high-$S_N$ phenomenon is rather a different matter.

In our view, and as discussed by \cite{whi87}, {\it any IGCs should be found
only in association with diffuse cluster light from intergalactic stars}.
That is, we consider it highly unlikely that globular clusters preferentially
condensed out of the hot, diffuse gas that pervades galaxy clusters,
leaving no other form of stellar material.  Furthermore, to generate a high
$S_N$ in the central galaxy, this IGC component must itself have (as we
will see quantitatively below) an extremely high and nonuniform $S_N$.
A plausible origin for this component of an intracluster medium could be from the
stripping of material from young, gas-rich galaxies {\it during} the collapse
of a cluster (see also \cite{mer84}), at an early enough stage that
many galaxies had formed globular clusters but had yet to
produce most of their halo field stars.  Most of the residual gas in these
young galaxies would then have been heated to the virial temperature
of the surrounding cluster and joined the ICM
which, in large clusters, outweighs the stellar material in the galaxies
(\cite{dfj90}; \cite{arn92}; \cite{whi93a}). 
This option is worth investigating in some detail, 
since it stands potentially to address, at once, the issues of BCG
specific frequency, diffuse light in galaxy clusters, and cD envelope
formation.

The first problem is to account for a correlation between cluster
X-ray luminosity and the specific frequency of the intergalactic stellar
material (and hence for the correlations of West \etal\ and B97).  Let 
us define a globular cluster formation efficiency
\begin{equation}
\epsilon \equiv {{N_{GC}}\over{M_{stars}+M_{gas}}} \ ,
\label{eq7}
\end{equation}
where $M_{stars}$ is the mass in the stellar component of the galaxy
and $M_{gas}$ is the mass of the leftover gas unused for star formation.
We can then write the presently observed specific frequency as 
\begin{equation} S_N\propto {{N_{GC}}\over{M_{stars}}} = 
\epsilon\ \left(1+{{M_{gas}}\over{M_{stars}}}\right)\ .
\label{eq8}
\end{equation}
Suppose now that $\epsilon$ is basically similar from place to place (see HP94;
we will also use this assumption in the next section). Then, 
any galaxies in which star formation proceeded nearly to completion
($M_{gas}/M_{stars} \ll 1$) would end up with essentially equal (and
normal) $S_N$ set by the value of $\epsilon$. However,
many slow-forming, late-type galaxies
could well have been destroyed at a point in their
lives when $M_{gas}$ was still large. 
The stars and globular clusters from these shredded
gas-rich objects would also have spilled into the cluster at large, 
producing an $S_N$ larger than normal by a factor $\sim (1 + M_{gas}/ M_{stars})$.
In absolute terms, we could suppose that the collapse of more 
massive galaxy clusters engendered the destruction of larger numbers 
of gaseous systems, thus leaving behind larger ratios of 
gas mass to total stellar mass, as is observed (\cite{dfj90}; \cite{arn92}).
As well, this effect could contribute to the growth of cD envelopes and the
establishment of a correlation between cD envelope luminosity and the X-ray
luminosity $L_X$ of the surrounding cluster gas (\cite{sch88}; \cite{lop97}).

Pursuing this option, let us now assume that
a population of IGCs with specific frequency $S_N^{IGC}$, and an envelope
composed of diffuse intracluster light $L_{env}$, are
superimposed on a BCG body with normal $S_N^{body}$. The
apparent specific frequency of the combination will be 
\begin{equation}
S_N^{tot}={{S_N^{body}+(L_{env}/L_{body})S_N^{IGC}}
\over{1+L_{env}/L_{body}}} \ .
\label{eq9}
\end{equation}
Since $S_N^{IGC} > S_N^{body}$ by hypothesis, the addition of the IGC
component will increase $S_N^{tot}$.
$L_{env}/L_{body}$ increases weakly with the parent cluster
$L_X$ (\cite{sch88}).  Thus, $S_N^{tot}$ would increase with $L_X$ 
even if $S_N^{IGC}$ were constant from cluster to cluster.  
However, $S_N^{tot}$ in cD galaxies should then also correlate with the
relative size of their cD envelope, $L_{env}/L_{body}$.  This is {\it
not} an observed trend (for example, M87 itself has a rather modest envelope
but one of the highest and best-measured specific frequencies; and other
BCGs with very massive envelopes such as NGC 6166 have unexceptional $S_N$
values; see the next section).  

The alternative is to let $S_N^{IGC}$ also 
increase with cluster $L_X$ (or mass, or velocity dispersion, or
temperature). By Eq.~(\ref{eq8}), this will occur if the formation of more
massive galaxy clusters somehow involves the destruction of
protogalaxies with systematically larger gas fractions $M_{gas}/M_{stars}$.
In this case, the initial metallicity of the
intracluster gas should be {\it anti}-correlated with $L_X$ (see also
\cite{dfj90}). Even though some subsequent contamination due to galactic winds
from the surviving protogalaxies must certainly have occurred (\eg,
\cite{mat95}), such an anti-correlation does appear to exist (\cite{fab94}).
Thus, this scenario would allow IGCs to resolve at least some aspects 
of the high-$S_N$ problem if we allow $S_N^{IGC}$ to increase
with cluster mass. Unfortunately, this descriptive scenario cannot 
account {\it quantitatively} for other features of the GCS,
as we will see next.  

Let us try to construct a model for M87 specifically, by building it from 
an M49-like body with a normal GCS, surrounded by a much more extended IGC envelope.
In Fig.~\ref{fig14} we model the observed $B$-band surface
photometry of M87 (\cite{car78}) as the sum of two components. One is the
underlying body of the galaxy, with a conventional $r^{1/4}$ law profile,
\begin{equation}
\mu_B^{body}=14.615 + 2.658\,(r/{\rm arcsec})^{1/4}\ 
{\rm mag\ arcsec}^{-2}\ .
\label{eq10}
\end{equation}
The other is a faint envelope which, for the sake of our argument, is assumed
to be part of the intracluster medium around M87:
\begin{equation}
\mu_B^{env}=26.675+1.875\, \log\, [1+(r/1200\arcsec)^2]\ 
{\rm mag\ arcsec}^{-2}  .
\label{eq11}
\end{equation}
The envelope necessarily has a very large core radius of
$24\farcm 7\simeq 108(D/15)$ kpc.  Beyond this core radius, the envelope
light behaves roughly as $I_{env}\propto r^{-1.5}$.  
We note that Carter \& Dixon (1978) find an especially bright
and extensive envelope component by comparison with
de Vaucouleurs \& Nieto (1978) or other authors; thus, our two-component
model is a ``maximal-envelope'' solution.  

Integration of Eqs.~(13) and (14) shows that in the radial
regime $1\arcmin\ltsim r\ltsim 6\farcm 5$, over which Cohen \& Ryzhov
(1997) have recently obtained high-quality radial velocities for 205 globular
cluster candidates in M87, the envelope contributes to the total light of the
galaxy in the ratio $L_{env}/L_{body}\sim0.10-0.15$.
The total specific frequency of M87, averaged over this same
interval, is roughly $S_N({\rm M87})\sim 11$,
while that of M49 (which we shall take to represent the 
underlying $S_N^{body}$ of M87) is $\sim 4.7$. Thus
the fraction of the globulars around M87
that are actually bound to it would be
\begin{equation}
{{N_{body}}\over{N_{tot}}} = {{S_N^{body}}\over{S_N^{tot}}}\,
\left(1+{{L_{env}}\over{L_{body}}}\right)^{-1} \sim 0.4\ ,
\ \ \ 1\arcmin \ltsim r \ltsim 6\farcm 5\ .
\label{eq12}
\end{equation}
The globulars bound to the galaxy will have a Gaussian velocity distribution 
with an intrinsic dispersion $\sigma_v\sim 270$ km s$^{-1}$ characteristic of
the stellar core of the galaxy (\eg\ \cite{sem96}), while the IGCs that are
spatially superimposed will have $\sigma_v\sim 570$ km s$^{-1}$ typical of
the early-type galaxies in Virgo (\cite{bts87}).  
The combination of the two, in the proportions given by Eq.~(\ref{eq12}), 
can be reasonably well described by a Gaussian with $\sigma_v\sim 400$ km s$^{-1}$,
which is not inconsistent with the data of Cohen \& Ryzhov (1997). 
However, the most noticeable effect of any IGCs would show up
in the wings of the velocity distribution:  the Cohen-Ryzhov sample
of 205 globulars built in these proportions should have $\sim 5$ clusters
with $v>1000$ km s$^{-1}$, whereas the actual sample has none.

A still more stringent test comes from
the decomposition of the M87 surface brightness, together with the observed
specific frequency, into putative ``body'' and ``envelope'' (IGC) components.
Suppose that the body of M87 has a specific frequency profile
$S_N^{body}$ identical to that of our `template' galaxy M49 (shown in 
Fig.~\ref{fig15}, along with 
the total $S_N^{tot}$ observed for M87).  
{\it By hypothesis}, the difference between $S_N^{tot}$ and $S_N^{body}$ 
is due to the intergalactic envelope, which allows us to find
$S_N^{IGC}$ as a function of $r$.
The results of this exercise are shown in Fig.~\ref{fig15}b
as the solid dots. The open squares in the same figure show
the $S_N^{IGC}$ that is inferred if the ``body'' of M87 is assumed instead
to have $S_N = 5$ throughout, with no dependence on galactocentric
radius.

The implications of this exercise are severe. In particular, we find that
$S_N^{IGC}\propto r^{-1\pm0.25}$ between $r\simeq 1\farcm 2$
and $r\simeq 9\farcm$ ($\simeq 5$ to 40 kpc). Because the cD
envelope, or diffuse cluster light, has an essentially constant surface
brightness at these radii (see Fig.~\ref{fig14}), this means that $S_N^{IGC}$ is
directly proportional to the projected number density $\sigma_{IGC}$ of the
supposed intergalactic globulars; thus, we conclude that
$\sigma_{IGC}\propto R_{gc}^{-1\pm0.25}$. That is, 
we are forced to claim that the intergalactic clusters are significantly
concentrated towards the center of the galaxy itself, {\it with a core radius no
larger than $\sim 1\arcmin=4.3$ kpc} (that is, the same core radius as the
M87 GCS itself). Such behavior is unacceptable for
any collection of truly intergalactic objects, and completely contradictory to
our original hypothesis. It is also worth noting that the deduced absolute value
of $S_N^{IGC}$ reaches a level in the core ($> 100$) 
that is an order of magnitude larger than in any known galaxy.

We might try to get around the strong radial dependence
of the IGC cluster component 
by assuming an $S_N^{IGC}$ that is independent of galactocentric radius,
and then use this and the observed $S_N^{tot}$ and $S_N^{body}$ to solve 
for the ratio $L_{env}/L_{body}$. The observed total
surface-brightness profile of M87 
can then be used to find the envelope intensity $I_{env}$ as a function of
$r$. However, this rather desperate argument
results in a diffuse light component that varies
essentially as $I_{env}\propto r^{-1}$ down to $\sim 1\arcmin$ distances
from the center of M87 -- which again implies that this ``intergalactic''
material must in fact be bound to the galaxy.  Finally, we note also that if we had
used the de Vaucouleurs \& Nieto (1978) photometry instead, the solution for the
envelope (Fig.~\ref{fig14}) would have been considerably fainter,
$S_N^{IGC}$ even larger, and thus even less favorable to the scenario.

It is important to emphasize more descriptively why the IGC model fails.
The key is that the ``specific frequency problem'' embodied by M87
is a {\it local} problem everywhere in the halo   
as well as a {\it global} one.  M87 has an
anomalously large cluster population at all galactocentric radii $r$
and not just in its cD envelope.  Thus, for example, we cannot turn M49
into M87 simply by adding large numbers of extra clusters to its
outskirts; they must be added in the same proportions all the way in
to the inner halo and core of the galaxy as well (\cite{har86}).  
This requirement then
forces the postulated IGC profile to have a central concentration
resembling the underlying galaxy, as well as an inordinately high specific
frequency.  There appears to be no natural 
way to do this through the IGC model, because
a realistic IGC component, or a population of clusters tidally stripped and 
accreted from other galaxies, would preferentially
boost the cluster population in the outer halo of the galaxy while
leaving the inner halo relatively unchanged.  

The argument we have just outlined may, however, not apply to all
high-$S_N$ BCGs.  Some other BCGs have relatively much larger cD
envelopes and thus, probably, stronger contributions from the
central cluster potential well.  
The intriguing case of NGC 1399, the BCG in the Fornax cluster, may
represent one such situation.  Recent kinematic studies of its GCS
(\cite{kis97a}; \cite{min97}) show that for radii $\gtsim 20$ kpc
the velocity dispersion of its globular clusters matches the 
surrounding Fornax galaxies rather than the NGC 1399 halo light 
(a difference of a factor of two in $\sigma(v_r)$).  We are clearly
just beginning to understand how these rather unusual galaxies
were built.

\subsection{A Formation Scenario}

The preceding arguments favor the view
that the essential features of the M87 GCS, and the other
high-$S_N$ cD systems like it, were built in at very early times.  Somewhat by
default, this ``initial conditions'' route has usually relied on the
supposition that the {\it efficiency of globular cluster formation
$\epsilon$} as defined in Eq.~(\ref{eq7}) could have differed  
by about one order of magnitude {\it ab initio} among E, cD, dwarf, 
and spiral protogalaxies, thus directly creating the large
range in $S_N$ that we now see (\eg, Harris 1981, 1991; \cite{vdb84};
HP94).  It is possible that the amount of gas cloud shocking and turbulence
could have directly governed $\epsilon$, driving it up to higher values in
denser or more massive protogalactic environments where collision speeds were higher.
Unfortunately, to date this has been a somewhat {\it ad hoc} view with
little observational evidence that
could be brought to bear one way or the other,
and so it, too, has been a less than satisfactory option.

The recent work of B97 and BTM97 opens the door to an alternate
approach to this problem.  Their data (see above)
imply more clearly than before (see
\cite{wes93} and \cite{kum93} for similar earlier suggestions) that $S_N$
increases with the overall mass density of the protogalactic 
environment.  B97 proposes essentially
that the high-$S_N$ galaxies do not have excess numbers of clusters; instead,
they have artificially low total luminosities relative to their globular
cluster populations.  In this picture, the ``missing
light'' is in the hot intracluster gas that never formed stars.

In such a scenario, then it is natural to ask why
this gas was unused for star formation, and how it ended up where it did.
The earliest $\sim 1$ Gyr history of an E galaxy is expected to
be heavily influenced by the SNII supernova rate from the first generation
of massive stars, and the subsequent development of a galactic wind
(\cite{mat71}; \cite{lar74}; see \cite{loe96} or Gibson 1994, 1996, 1997
for recent overviews).
Under the right conditions a large fraction of the original protogalactic 
gas can be ejected outward by the wind to join the ICM.
It seems highly likely to us that galactic winds 
would have influenced the GCS specific frequency, and by different degrees
in different galaxies.  Our reason for proposing this is based 
primarily on the timing of the GCS formation epoch:
a large protogalaxy should accumulate from
many smaller ($10^8 - 10^9 M_{\odot}$) dwarf-sized
gas clouds (supergiant molecular clouds or SGMCs), which provide 
the right host environments for globular
cluster production (\cite{sea78}; Larson 1990b, 1993; HP94).  
In this picture, we postulate that each SGMC contains several 
{\it protocluster cores}, which are the much denser sites where
star formation can take place at the necessary $\gtsim 50$\%
efficiency to produce a bound star cluster.  Since the protocluster
cores are expected to build up over timescales $\sim 10^8$ y
out of the gas in SGMCs (HP94; \cite{mp96}; \cite{ee97}),
it is likely that they would form very early in the evolution of
the protogalaxy, before the lower-density majority of the gas
began star formation in earnest.
Thus, the globular clusters should already be present before the wind
reaches its peak at a few $\times 10^8$ y
and interrupts subsequent field-star formation.

In essence, we suggest that {\it for normal E
galaxies} with specific frequencies $S_N \ltsim 5$,
the early galactic wind did {\it not} eject large amounts of the
original gas supply.  Today, these galaxies 
have GCS specific frequencies 
fairly close to the original cluster formation efficiency.  However, in
systems like M87, we propose
that the galactic wind was stronger, ejecting
much of the original gas mass outward into the dark-matter potential
well of the surrounding cluster, and leaving behind a GCS 
with an artificially high $S_N$.  

Is this idea a realistic one for gas within such massive potential wells as
we find in the BCGs?  (Normally, we would expect the {\it less} massive 
protogalaxies to be able to eject higher proportions of their initial gas.)
\cite{gib97} provides a comprehensive discussion of 
the way that various parameters 
affect the SNII production and mass ejection rate.  
Effective ways to increase the
amount of mass loss are (a) to increase the star
formation efficiency factor $\nu = \dot M_{\star}/M_{tot}$, which would
boost the early supernova rate; or (b) to decrease the rate 
at which the expanding supernova remnants can cool radiatively and lose energy;
or (c) to adopt an IMF flatter than the normal
Salpeter ($x = 1.35$) slope, thus increasing the number of Type II SNe per
unit mass.  Other studies have suggested that an IMF slope $x \simeq 1$
is also appropriate for matching the
observed Fe and O abundances in the ICM gas (\eg\, \cite{loe96}; \cite{ari87}; 
\cite{ren93}; \cite{dav91}, among others).  We are, however, looking for
a way to generate a {\it difference} in the mass loss rate (and hence $S_N$) 
in a BCG like M87, compared with normal systems like M49.  Given that these galaxies have
rather similar compositions and structures, we see no compelling reasons to
suggest arbitrary differences in the SNR cooling rates, or the IMF.
Thus of the various options mentioned above, the most plausible appears to us to
be the star formation rate $\nu$.  Our scenario therefore boils down to the
suggestion that BCGs had an extraordinarily high $\nu$ at early times.

Another important point to consider is that the simple
one-zone wind models mentioned above are an oversimplification 
for large protogalaxies, which were likely to have been highly clumpy.
Thus the parametrizations of mass loss rates 
for monolithic models such as the ones reviewed
by \cite{gib97} should not be taken too literally.
For a central supergiant like M87,
which is at the gravitational center of its surroundings, 
the protogalaxy would have been the site of a considerable amount of gas 
infall and chaotic, violent motion which could
have encouraged early, rapid star formation and a more massive SNII-driven wind.
This picture is consistent with the considerable evidence that many large
E galaxies almost certainly formed at epochs predating $z \sim 3$,
and some probably began major star formation at $z \gtsim 5$
(\eg, \cite{lar90a}; \cite{mao90}; \cite{tur91}; 
\cite{whi93a}; \cite{loe96}; \cite{mus97};
\cite{ste96}; \cite{gia96}; \cite{ben96}; \cite{ell97}, to name a few;
see also HP94 for a timing argument in support of a
redshift $z \gtsim 5$ for the first epoch of globular cluster formation).

We emphasize again that we are not suggesting this approach as the way
for a {\it typical} galaxy (elliptical or spiral) to form.  
The BCGs are, clearly, rather unusual entities, and the formation processes
during the much higher-density regime of $z \gtsim 5$ are entirely likely
to have been different from those at $z \ltsim 2$ which gave rise to the
main population of galaxies that we now see.
A fully realistic model of a BCG-type protogalaxy during the epoch of 
globular cluster formation and galactic wind outflow, in all
its undoubted complexity, is far beyond the
scope of this paper (or indeed any current models).  However,
we can roughly quantify the effect on the specific frequency as follows:
from Eqs.~(\ref{eq4}), (\ref{eq7}), and (\ref{eq8}), we obtain
\begin{equation}
S_N = 85.5 \times 10^6 \left(M \over L \right)_V \,\epsilon\,
\left(1+{{M_{gas}}\over{M_{stars}}}\right)
\label{eq14}
\end{equation}
where $(M/L)_V$ is the mass-to-light ratio for the stellar material
in an E galaxy; we adopt $(M/L)_V = 8$ (\cite{fab78}; \cite{bin87}).
$M_{gas}$ is the amount of gas assumed to have been driven out
by the galactic wind and unused for star formation.
With the explicit assumption $\epsilon \simeq const$, 
Eq.~(\ref{eq14}) can be rewritten as 
\begin{equation}
S_N = S_N^0 \,\left(1+{{M_{gas}}\over{M_{stars}}}\right)
\label{eq15}
\end{equation}
where $S_N^0$ represents the fiducial specific frequency for a galaxy in
which little or no gas is ejected and the star formation runs to
completion ($M_{gas} \ll M_{stars})$.  Lacking a full physical theory, 
we must estimate $S_N^0$ on observational grounds:

\noindent (a) The mean specific frequency for 14 large ellipticals in
Virgo, Fornax, and the NGC 5846 group (from data in Harris \& Harris 1997;
see also \cite{har91}; \cite{kis97}) is 
$\langle S_N \rangle = 5.4 \pm 0.3$, excluding the
cD's M87 and NGC 1399.  Since even the normal ellipticals have probably
lost some of their original gas to early winds, this value should
give an upper limit to $S_N^0$.  An extreme {\it lower} limit can be set
at $S_N^0 \sim 2$, which is the mean for the lowest specific frequencies
found in a range of E and dE galaxies in sparse groups.
A straight mean over {\it all} the non-BCG galaxies, weighted inversely as the
observational uncertainty in $S_N$, gives $\langle S_N \rangle = 3.3 \pm 0.2$.

\noindent (b) The mean specific frequency for 13 dwarf ellipticals
more luminous than $M_V^T \simeq -15$ (\cite{dur96}) is
$\langle S_N \rangle = 4.2 \pm 0.5$.  These same dE's are small 
enough to have lost typically $\gtsim 20$\% of their initial gas 
in a wind, according to published models
(\eg\ Gibson 1997).\footnote{Among dwarf E galaxies 
with globular cluster systems of their own, 
the lowest-luminosity dE's have the highest specific
frequencies (\cite{dur96}), the most extreme cases being the
Fornax and Sagittarius dwarfs with $S_N \sim 25$.  These tiny systems are likely
to be ones which expelled most of their original gas 
in a single early burst (\cite{dek86}).}
However, they may also have later destroyed a significant fraction of their
original number of globular clusters by dynamical friction, since the
GCS in a dwarf elliptical is quite centrally concentrated (see the
discussion of Durrell \etal).  These two effects have opposite 
and comparable influences on $S_N$, and it is not immediately clear 
which will be the more important over a Hubble time.

Taking the preceding arguments to bracket our estimate of $S_N^0$, we adopt
$S_N^0 = 3.5 \pm 1$.  (Similar `baseline' values for $S_N^0$ were
reached by West \etal\ 1995 and BTM97, though for different reasons.)
From Eq.~(\ref{eq14}), we find that the corresponding
fiducial cluster formation efficiency is 
$\epsilon \simeq 5.8 \times 10^{-9} M_{\odot}^{-1}$, or 
one globular cluster per $1.7 \times 10^8 M_{\odot}$ of gas.
For a mean cluster mass $\simeq 3 \times 10^5 M_{\odot}$
(from HP94; \cite{har96}; \cite{mey97}), the corresponding {\it specific mass}
ratio as defined by HP94 is $S_M = M_{GCS}/M_{stars} = 1.8 \times 10^{-3}$.
This result agrees extremely well with the observations from contemporary
star-forming regions in the Milky Way and M31, in which the mass of gas
seen to be forming clusters in giant molecular clouds 
is about $2 \times 10^{-3}$ of the total GMC mass (HP94; \cite{mcl97}).

The value of $S_N^0$ indicates that
only 0.2\% of the stellar mass in a typical E galaxy today is
in the form of bound star clusters.
To obtain a case like M87 with $S_N \sim 14$, we must then require
$(M_{gas}/M_{stars}) \simeq 3$, \ie\ that 75\% of the
original gas supply in the protogalaxy was ejected to rejoin the ICM.
More generally, if $M_{gas}$ equals the amount expelled by the wind,
then the fraction $f_M$ of the {\it initial protogalactic gas} that is
lost is $f_M = M_{gas} / (M_{gas} + M_{stars}) = 1 - (S_N^0/S_N)$.
This quantity, for our BCG sample, is shown in Fig.~\ref{fig16}.
For comparison, the expected dependence of $f$ on $\sigma$ derived from
the scaling relations Eqs.~(7-9) above is shown by the dashed line.  A rough linear
relation $f_M \sim (\sigma_{cl}/{\rm 1200\, km\, s^{-1}})$ as shown in
the graph matches equally well, given the observational scatter.

From this graph, we see again that to save the assumption  
$\epsilon \sim$ const, we have to claim that quite a substantial
fraction of the initial gas mass was expelled from the biggest systems.
In terms of the total gas mass
in the ICM, this is not at all an excessive requirement, since large   
Abell-type clusters typically contain $\sim 10^{13} - 10^{14} M_{\odot}$ 
of diffuse gas, most of which is likely to be primordial material 
never processed through galaxies (\eg\ \cite{dav91}; \cite{whi93a}).  
We recognize, however, that this
requirement imposes very strong demands on the early SNII rate 
which will have to be put to the test in more advanced modelling.

Lastly, we explore the dependence of total GCS size on the {\it total}
core mass in the surrounding cluster including the dark matter.  
For an isothermal potential well, the central density is given by
$\rho_c = 9 \sigma^2/ 4 \pi G r_c^2$ where $r_c$ is the {\it projected} 
core radius that we directly observe.
The cluster core radii are known empirically (BTM97; \cite{sch93}) 
to scale approximately as 
$r_c \simeq 104\, {\rm kpc}\, (\sigma/{\rm 500 km~s^{-1}})^{0.6}$.
Thus we obtain
\begin{equation}
\rho_c \simeq 3.86 \times 10^{-3} M_{\odot}\,{\rm pc}^{-3} \left(\sigma_{cl} 
\over 500\, {\rm km~s}^{-1}\right)^{0.8} \, .
\label{eq19}
\end{equation}
and the total mass contained within the core is 
\begin{equation}
M_{core} \simeq {4 \over 3} \pi r_c^3 \rho_c = 
1.8 \times 10^{13} M_{\odot}\, \left(\sigma_{cl} 
\over 500\, {\rm km~s}^{-1}\right)^{2.6} \, .
\label{eq20}
\end{equation}
$M_{core}$ is of course
dominated by the dark matter, with smaller contributions from the visible
galaxy and the X-ray gas.  Since the core radii
are typically of order $\sim 100$ kpc, $\rho_c$ defined in this
way comfortably encloses the bulk 
of the central BCG and provides a reasonable
representation of the overall mass density there.
In Fig.~\ref{fig17} we show the ratio
$M_{core}/N_{GC}$, the {\it total mass per unit globular cluster},
plotted against velocity dispersion.  
The graph indicates that
the globular cluster formation efficiency {\it in an absolute sense}
(number per unit total mass, $M_{core} = M_{gas} + M_{stars} + M_{dark}$) 
is {\it lower} in the more massive BCGs.
This result is consistent with our suggestion that the galactic wind
may have been earlier and more effective in the more massive environments,
possibly interrupting even the GCS formation before it could run to completion.

This conclusion about the absolute efficiency of globular
cluster formation is not the same as in BTM97.  
This apparent disagreement is due primarily to different definitions of terms.
They defined an efficiency ratio using a {\it projected} core mass $M_c$ 
over a cylindrical volume, derived from the surface density $\Sigma_c$
rather than $\rho_c$.  In addition, they 
calculated $M_c$ for a fixed radius of 40 kpc in all clusters. 
We believe it is more
appropriate to use a total core mass calculated from a true three-dimensional
mass density, in order to compare with the total globular cluster population
in the entire BCG.  For the sake of more direct comparison, we show in
the lower panel of Fig.~\ref{fig17} the ratio $M^{100}/N_{GC}$, where
$M^{100}$ is the total mass within a fixed (spherical, not cylindrical)
volume of radius of 100 kpc, which roughly represents the total optical 
extent of these galaxies.  In the sense defined by $M^{100}$,
the absolute efficiency increases with $\sigma_{cl}$, though the
trend is less obvious than in the upper panel.  It is not clear to us
which of the two representations is preferable; in either graph,
a typical mass ratio is about $3 \times 10^9 M_{\odot}$ per globular cluster.
Comparing this value with the mean $\epsilon^{-1} \sim 2 \times 10^8 M_{\odot}$ 
(gas mass per globular cluster) derived above, we see that
the dark matter makes up $\gtsim 90$\% of the total mass even within
the cluster core.

The scenario we have outlined here explores the implications of 
a single important assumption:  that the conversion rate $\epsilon$ of 
protogalactic gas into globular clusters was statistically uniform
in different environments.  Pursuing this assumption 
leads to the hypothesis that BCGs ejected a much higher fraction of their
initial gas mass than did more normal E galaxies.
We recognize that this discussion is highly speculative,
and is supported only in an indirect sense by the GCS scaling relations
which are just now beginning to emerge from the data.  

The obvious alternative
to this admittedly risky conclusion is simply to assume that
$\epsilon$ is strongly dependent on environment, varying by up to an order
of magnitude in different E galaxies.  As yet, there are no definitive 
arguments either to rule out such a view or to clearly support it.
The galactic wind hypothesis is offered as an alternate approach,
which does has the advantage of connecting to a known major event 
in the early history of a large E galaxy.

\section{SUMMARY}

We have conducted a photometric study of the inner-halo globular cluster
system in the Virgo giant M87, and have used the characteristics of its 
GCS as the starting point for a discussion of the formation of 
central cluster galaxies like it.  A summary of our principal
findings is as follows:

\noindent (1) The luminosity distribution of the globular clusters in M87
(in its form either as the GCLF or the LDF) is virtually independent of
location in the halo for {\it any} projected radius $r \gtsim 1$ kpc.
We suggest that the main effects of dynamical evolution (tidal shocking
and evaporation) may be limited to clusters less massive than
$M_{cl} \ltsim 10^5 M_{\odot}$, below the faint limits of the present data.
Alternatively, cluster evolution has proceeded in such a way as to preserve
the overall shape of the GCLF as a function of time, as the recent
model simulations are beginning to show.
Dynamical friction may have removed the very most massive clusters
($M \gtsim 2 \times 10^6 M_{\odot}$) in the innermost few kiloparsecs.

\noindent (2) Within $\sim 4$ kpc of the galaxy center,
the mean metallicity of the GCS is uniform with radius  
at the metal-rich level of [Fe/H] $\simeq -0.3$.
At larger radii, the mean cluster metallicity declines as
$(Z/Z_{\odot}) \sim r^{-0.9}$.  We find that cluster metallicity is
largely uncorrelated with cluster luminosity.

\noindent (3) M87 holds the paradigmatic high-specific-frequency globular
cluster system.  We examine carefully the hypothesis that its many thousands
of ``extra'' globular clusters are due to an intergalactic (IGC) population
in the Virgo potential well, and reject it.  
The dominant quantitative problem with the
IGC model is that it requires the intracluster
globulars to have a highly unrealistic spatial distribution, inconsistent 
with the assumptions of
the model.  We conclude that the vast majority of the globular clusters
within M87 belong to it, and were formed with it {\it in situ}.

\noindent (4) Adding the recent results of Blakeslee (1997) and 
\cite{btm97} to other data from the literature, we reinforce the
the result that GCSs in ``brightest cluster galaxies'' like M87 become
continuously more populous (higher specific frequency) with 
parent galaxy size, and with the total mass of the surrounding galaxy
cluster.  However, the absolute {\it efficiency} of GCS formation
(measured as the number of globular clusters per unit mass in the cluster
core) {\it decreases} with increasing core mass.
We use the available observations for 30 BCGs to construct
empirical scaling relations describing the globular cluster population
and specific frequency as a function of cluster mass (measured by the
velocity dispersion $\sigma_{cl}$ of the galaxies).

\noindent (5) We suggest that globular cluster specific frequency in BCGs
may have been strongly influenced by the early galactic wind driven by the
first round of Type II supernovae throughout the protogalaxy.  In particular,
we speculate that BCGs differed from normal ellipticals by generating
a higher early star formation rate.  The considerably stronger resulting
wind interrupted subsequent star
formation, drove out a large fraction of the initial gas mass, and left behind
a galaxy with a higher than average ratio of globular clusters to field stars.
If this hypothesis proves to be invalid, the most obvious alternative is
to require that the GCS formation efficiency per unit gas mass ($\epsilon$) 
was higher in BCGs than in normal ellipticals, and in general becomes
progressively higher in larger-density environments.

\acknowledgments

This research was supported through grants to W.E.H.~and G.L.H.H.~from
the Natural Sciences and Engineering Research Council of Canada.
Additional support was provided to D.E.M. by NASA through grant
number HF-1097.01-97A awarded by the Space Telescope Science Institute,
which is operated by the Association of Universities for Research
in Astronomy, Inc., for NASA under contract NAS5-26555.

\clearpage

\begin{center}
{\bf Figure Captions}
\end{center}

\figcaption[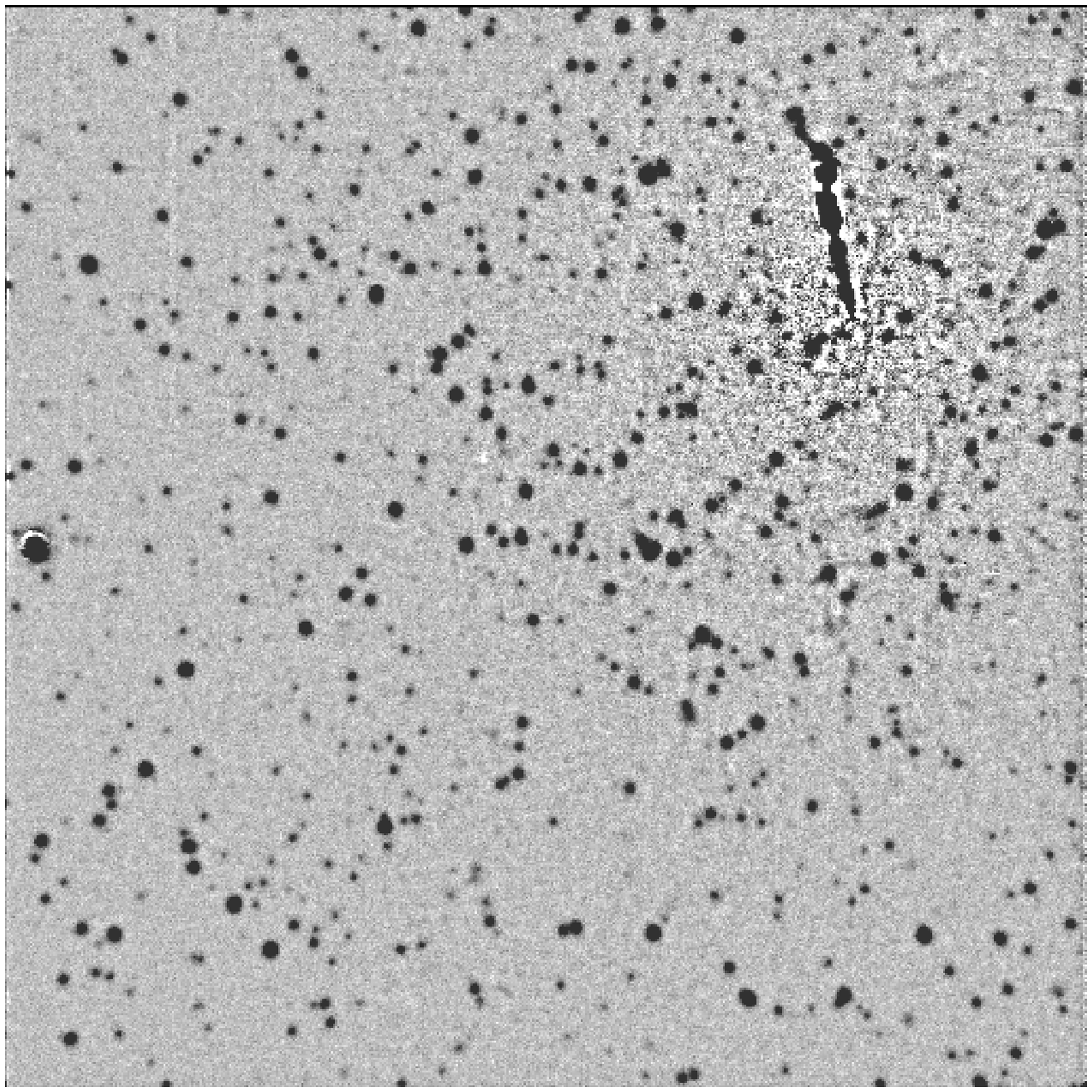]{$R-$band image of the central field in M87.  This
frame is a median of 26 individual 500-sec exposures in $R$ taken with the
CFHT and HRCam at prime focus (see Sec.~2.2 of the text).  North is approximately
at the upper right corner, East at lower right; see Fig.~2  for a finder
chart with an exact orientation map.  The large black dot along the left edge
of the field is the hole enclosing the bright guide star for the tip-tilt
correction of the camera.  The smooth isophotal contours of the galaxy have
been subtracted through an ellipse-fitting model (see Sec.~3).
\label{fig1}}

\figcaption[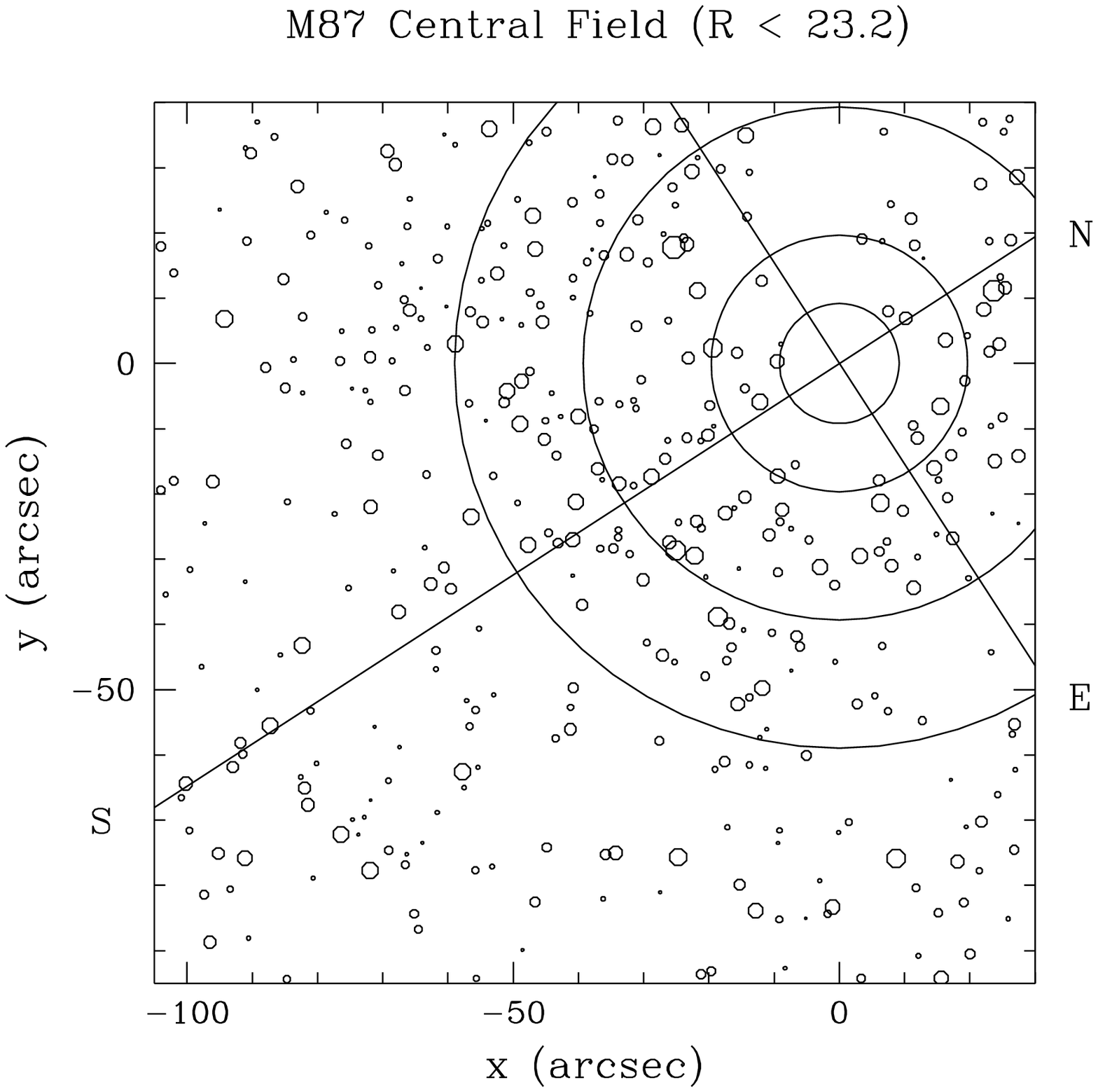]{Distribution of starlike objects in the M87
central field that are brighter than the GCLF turnover ($R \ltsim 23.2$).
Larger symbols correspond to brighter objects; compare with the direct image
of Fig.~1.  Fiducial directions are marked on the borders, with the galaxy
center at (0,0).  The concentric circles have successive radii of $9\farcs 2$,
$19\farcs 7$, $39\farcs 3$, and $59\farcs 0$, defining the four radial zones
used in the GCLF analysis.  Note the small sector directly above the galaxy
center with no measured objects:  this sector contains the nuclear jet and was
excluded from any analysis.
\label{fig2}}

\figcaption[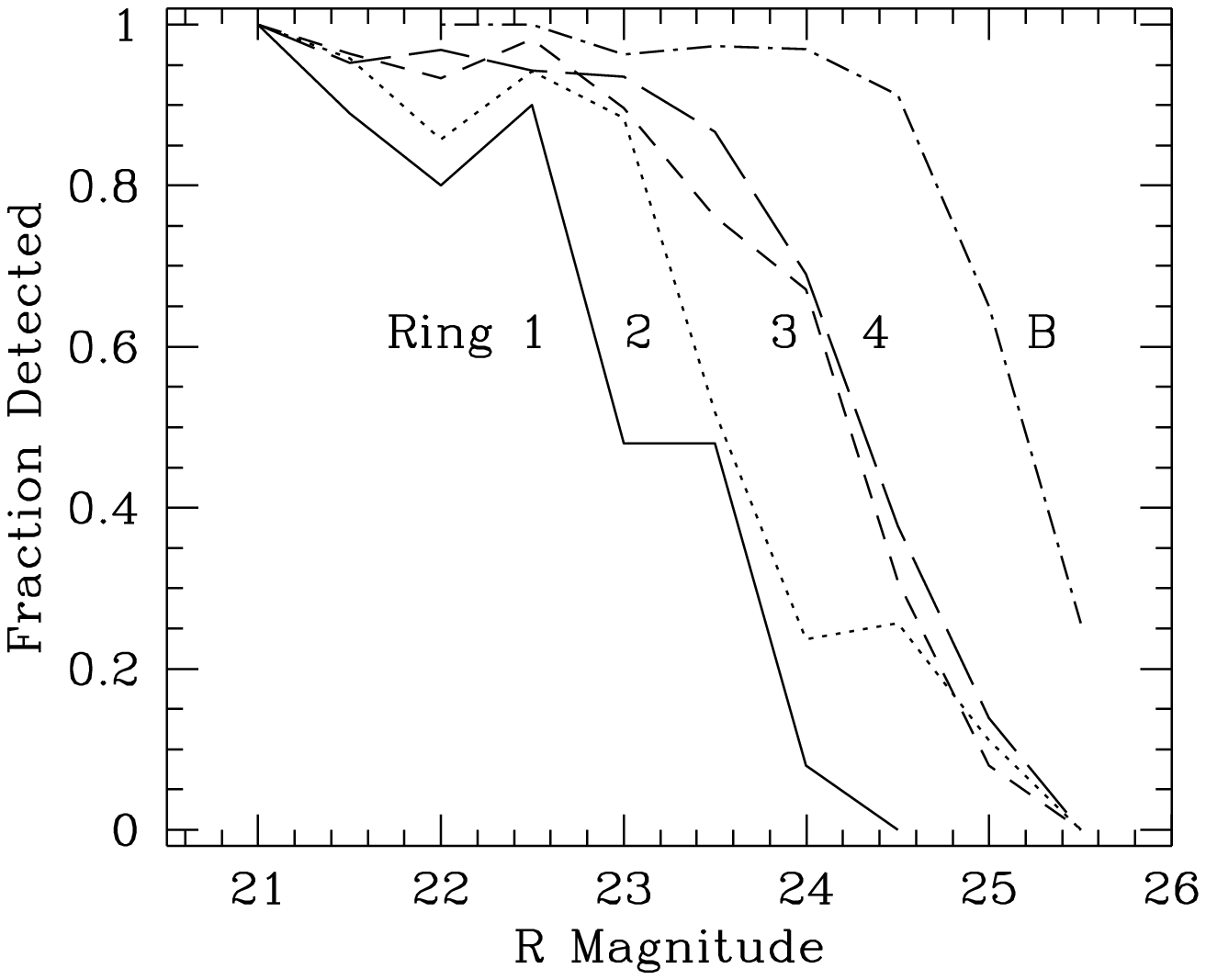]{Completeness of image detection in each annular
zone, as a function of $R$ magnitude.  These curves are the empirical results
of artificial-star tests as described in Sec.~2.2.  The five zones are defined in
Table 4:  Ring 1 is the innermost zone and B denotes the remote background
field.
\label{fig3}}

\figcaption[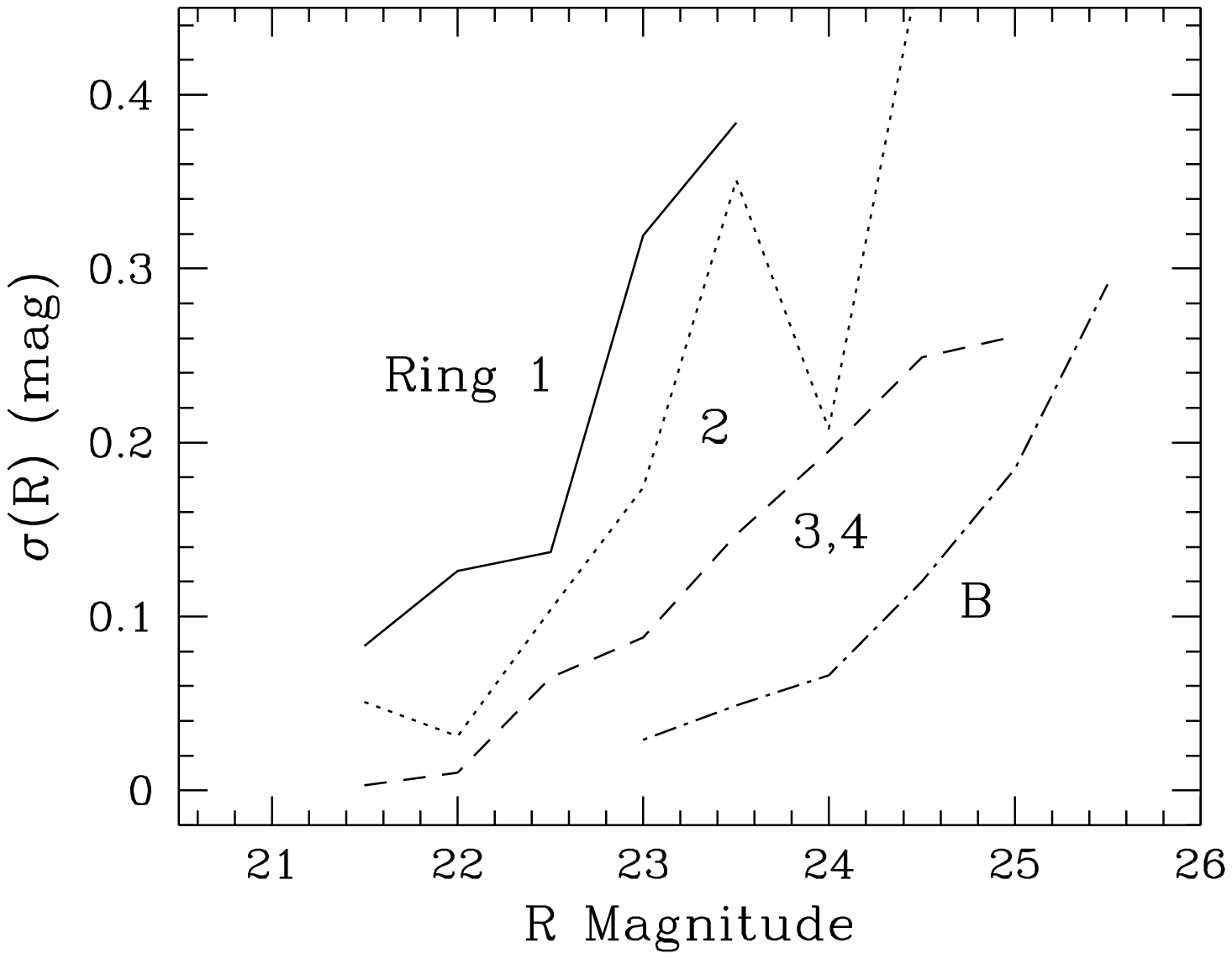]{Internal precision of the $R-$band photometry, as
derived from the artificial-star tests.  Here $\sigma(R)$ is rms scatter in
the measured magnitudes of the recovered stars.  The data are plotted
(unsmoothed) in 0.5-mag steps, with the five zones labelled as in the previous
figure.
\label{fig4}}

\figcaption[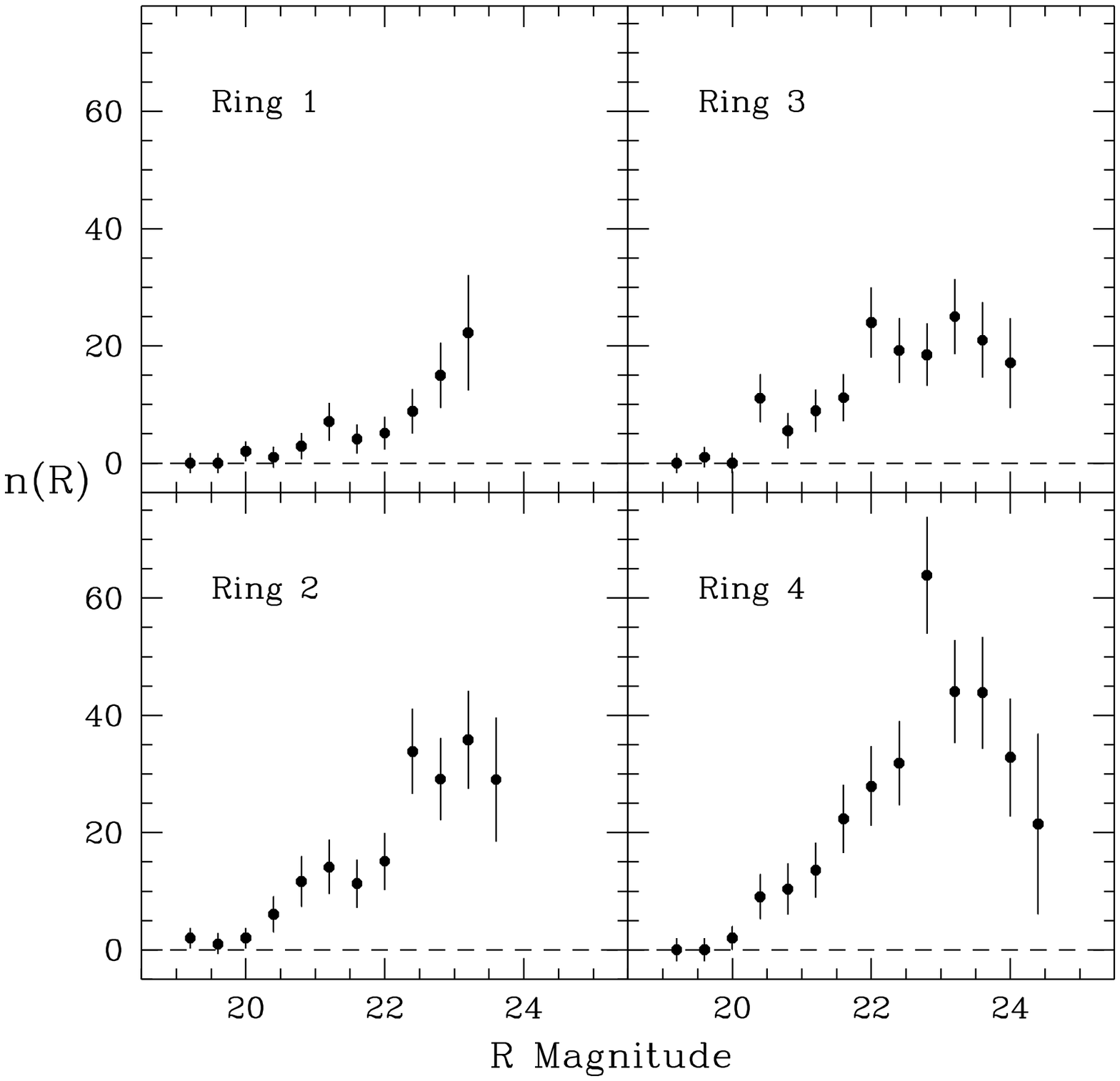]{GCLFs for the four zones defined in Table 4.  In
each bin, the number of globular clusters per 0.4-mag bin, fully corrected for
detection incompleteness and background subtraction, is plotted against $R$
magnitude.
\label{fig5}}

\figcaption[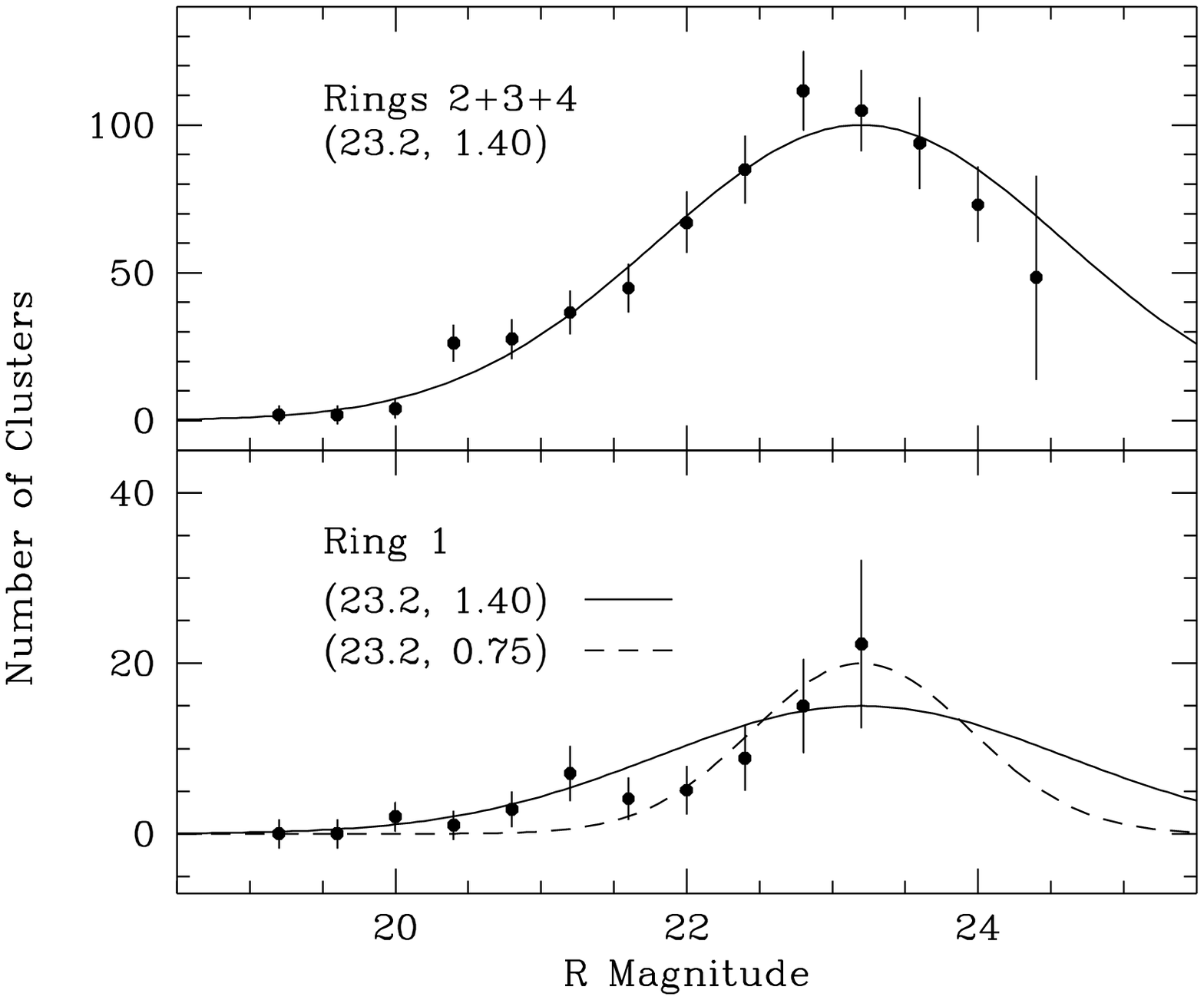]{(a) {\it Upper panel}:  The GCLF for the three
zones 2-4 combined.  A Gaussian curve with peak at $R_0 = 23.2$ and dispersion
$\sigma = 1.4$ mag as derived from Whitmore \etal\ (1995) and the present data
is superimposed on the datapoints.  (b) {\it Lower panel}:  The GCLF for the
innermost ring.  Two Gaussian curves with the same peak but different
dispersions are compared:  the narrower curve with $\sigma = 0.75$ provides
the formal best fit, but the wider one which matches the outer zones cannot be
strongly ruled out.
\label{fig6}}

\figcaption[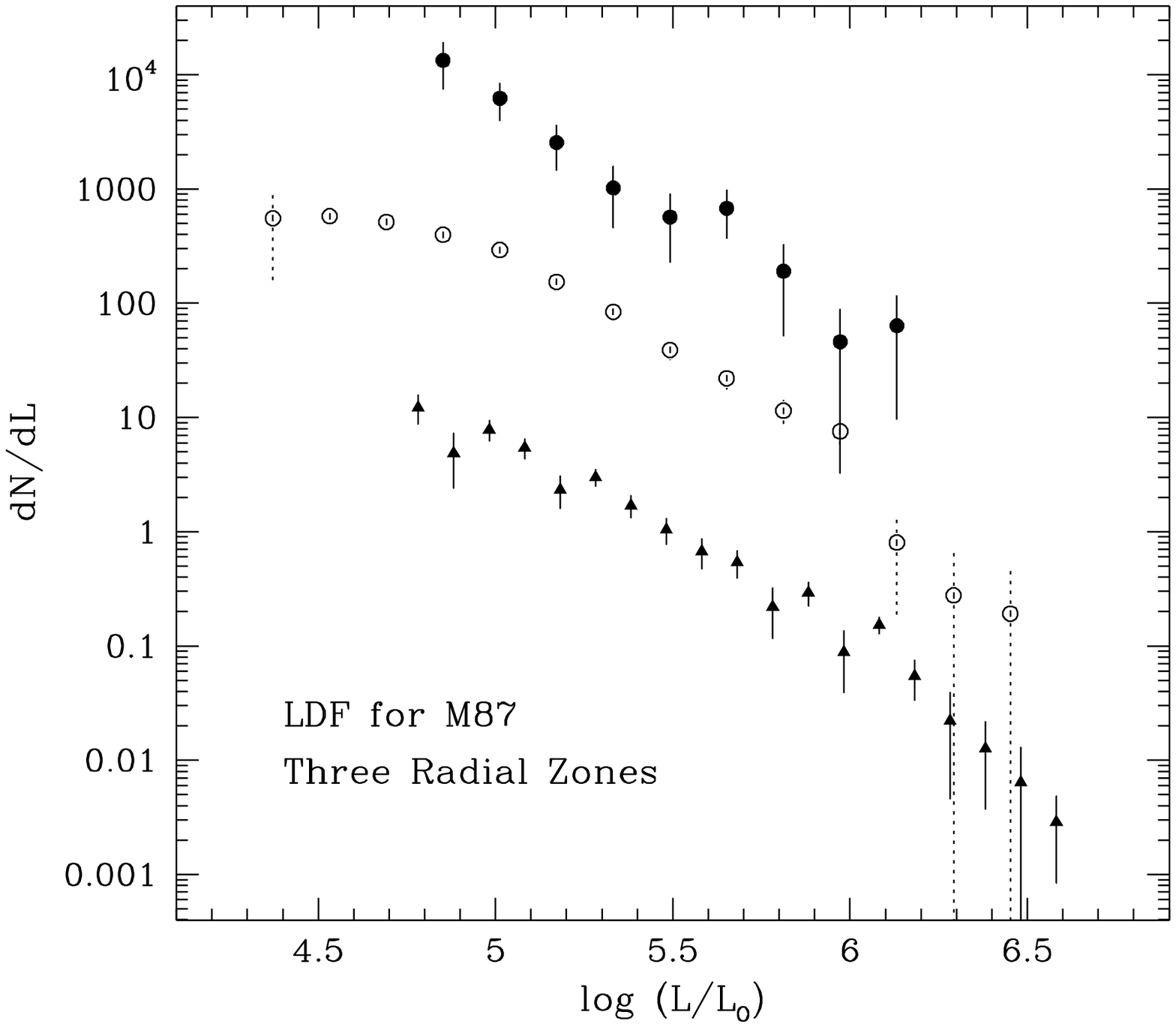]{Luminosity distribution function (LDF) for the M87
globular clusters in three regions of the halo.  Here $dN/dL$ represents the
number of clusters per unit luminosity, plotted against cluster luminosity in
Solar units $(L/L_{\odot})$; the three sets of points have been shifted
vertically by arbitrary amounts for clarity.  Solid circles 
are the data from ring 1 (projected radii $0.7 - 1.4$ kpc); 
open circles are from rings $2-4$ ($1.4 - 10.3$ kpc); and 
solid triangles are outer-halo GCS data ($r > 10$ kpc)
from \protect\cite{mhh94}.  Note the similarity of the LDF slope in all three
regimes over most of the luminosity range, coupled with a progressive
disappearance of the most luminous clusters (log $(L/L_{\odot}) \gtsim 6.2$)
going inward toward smaller radii.
\label{fig7}}

\figcaption[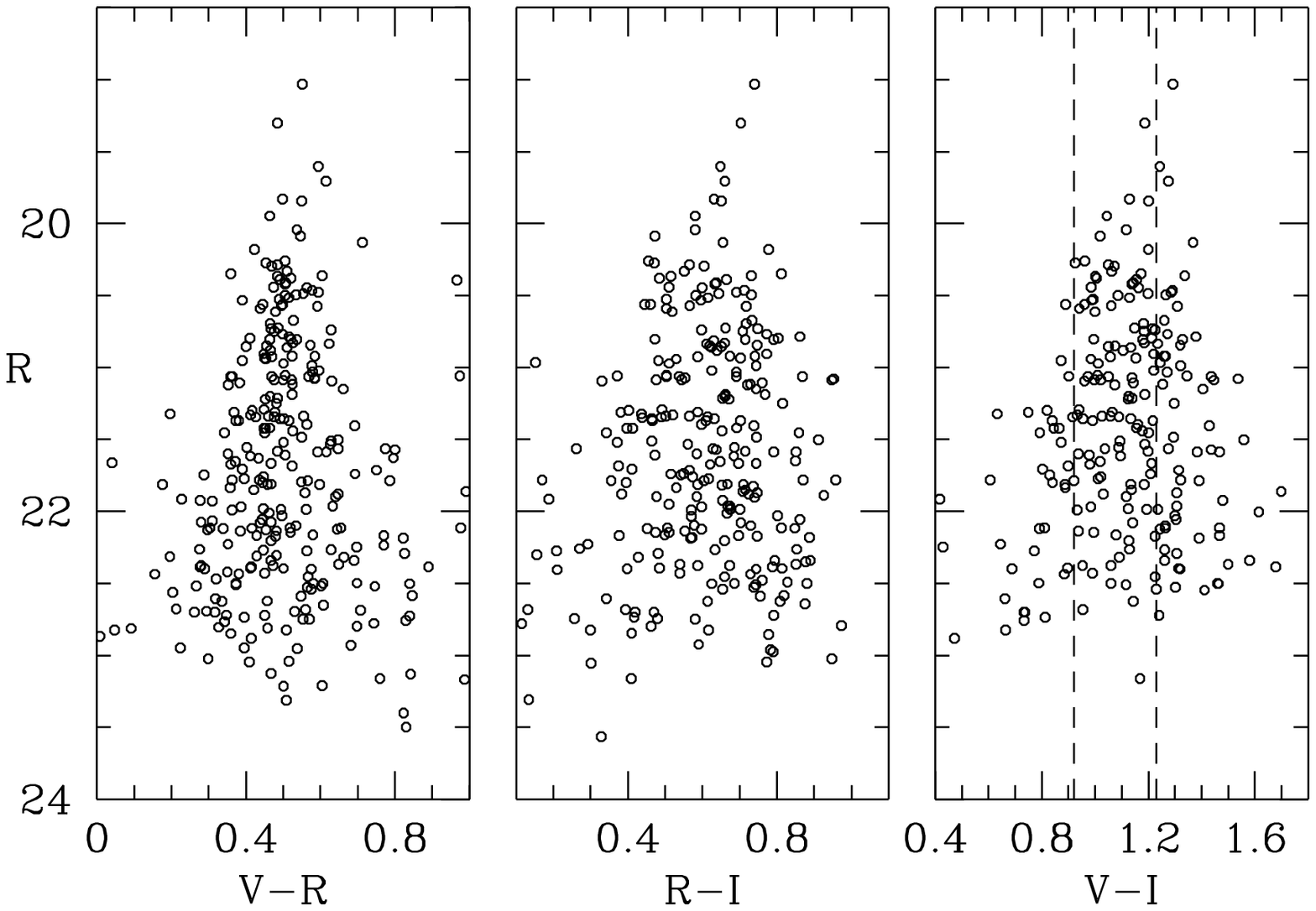]{Color-magnitude distribution for the measured
objects in the M87 central field; $R$ is plotted against $(V-R)$, $(R-I)$, and
$(V-I)$.  In the last panel, the two peaks in the bimodal color distribution
noted by \protect\cite{els96b} and Whitmore \etal\ (1995) are indicated by
dashed lines.  Most of our objects belong to the redder, more metal-rich group
since they are in the inner halo; see text.
\label{fig8}}

\figcaption[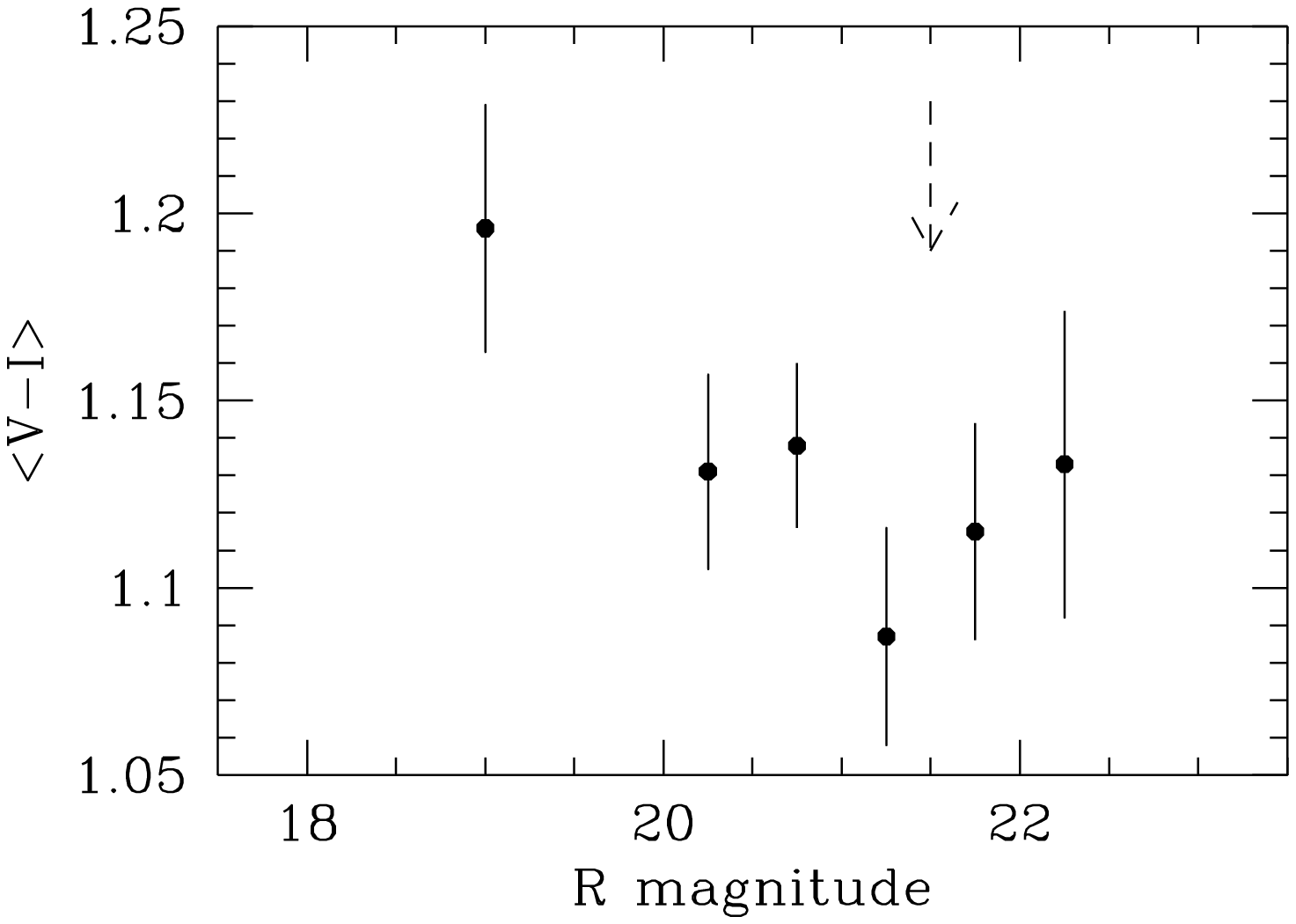]{Mean $(V-I)$ colors of the globular clusters as a
function of magnitude.  The dashed arrow at $R \sim 21.5$ marks the magnitude
near which \protect\cite{els96b} found an extra ``clump'' of clusters in the
blue part of the bimodal color distribution.
\label{fig9}}

\figcaption[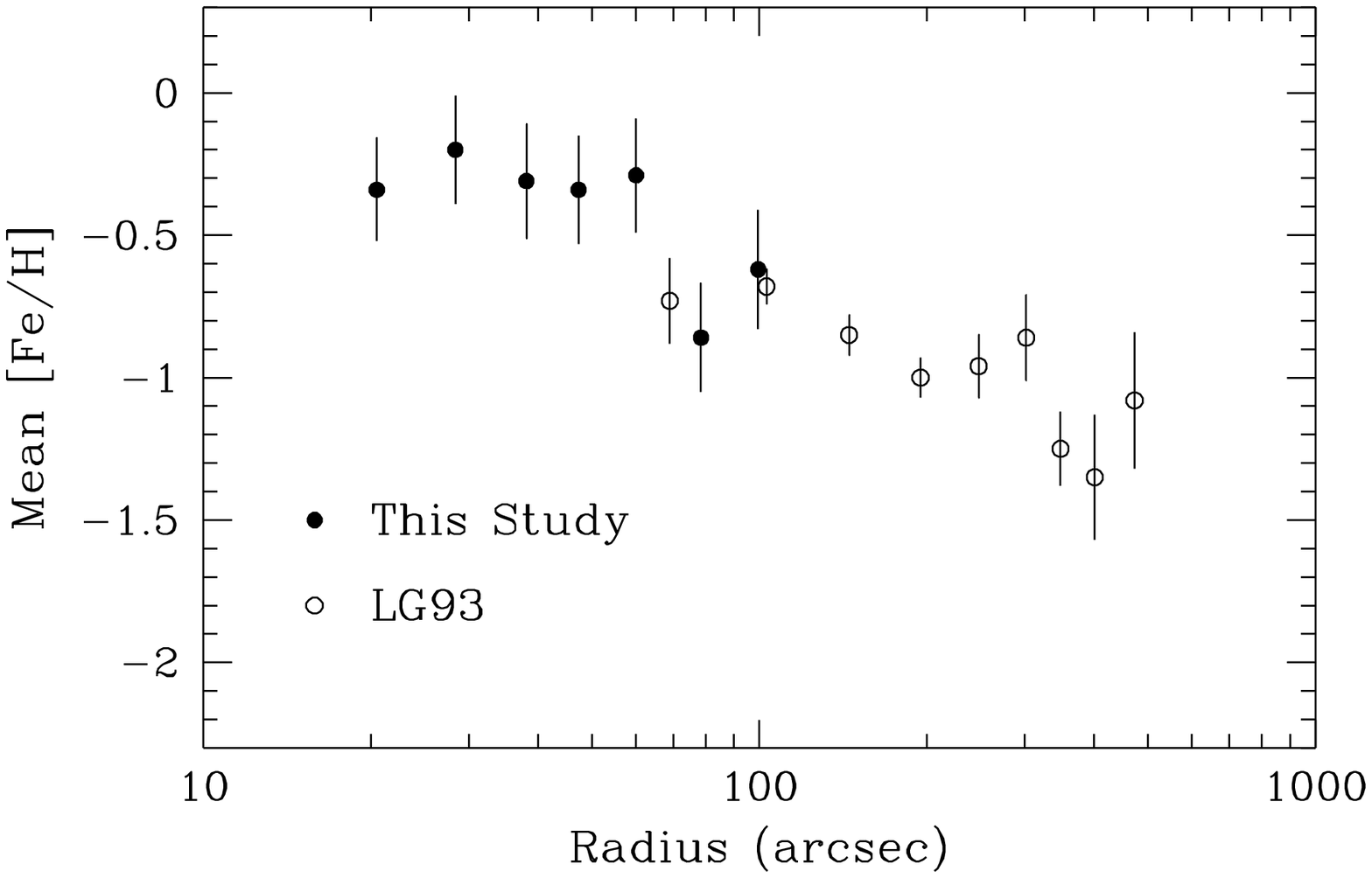]{Metallicity gradient for the globular cluster
system in M87, combining the inner-halo data from this study (solid symbols)
and the outer-halo data from \protect\cite{lee93} (open symbols).  Note that
within one core radius $r_c \simeq 1'$, there is no change in the mean cluster
metallicity; at larger radii, there is a steady decrease in mean metallicity
out to the limits of the observations, scaling roughly as $Z \sim r^{-0.9}$.
\label{fig10}}

\figcaption[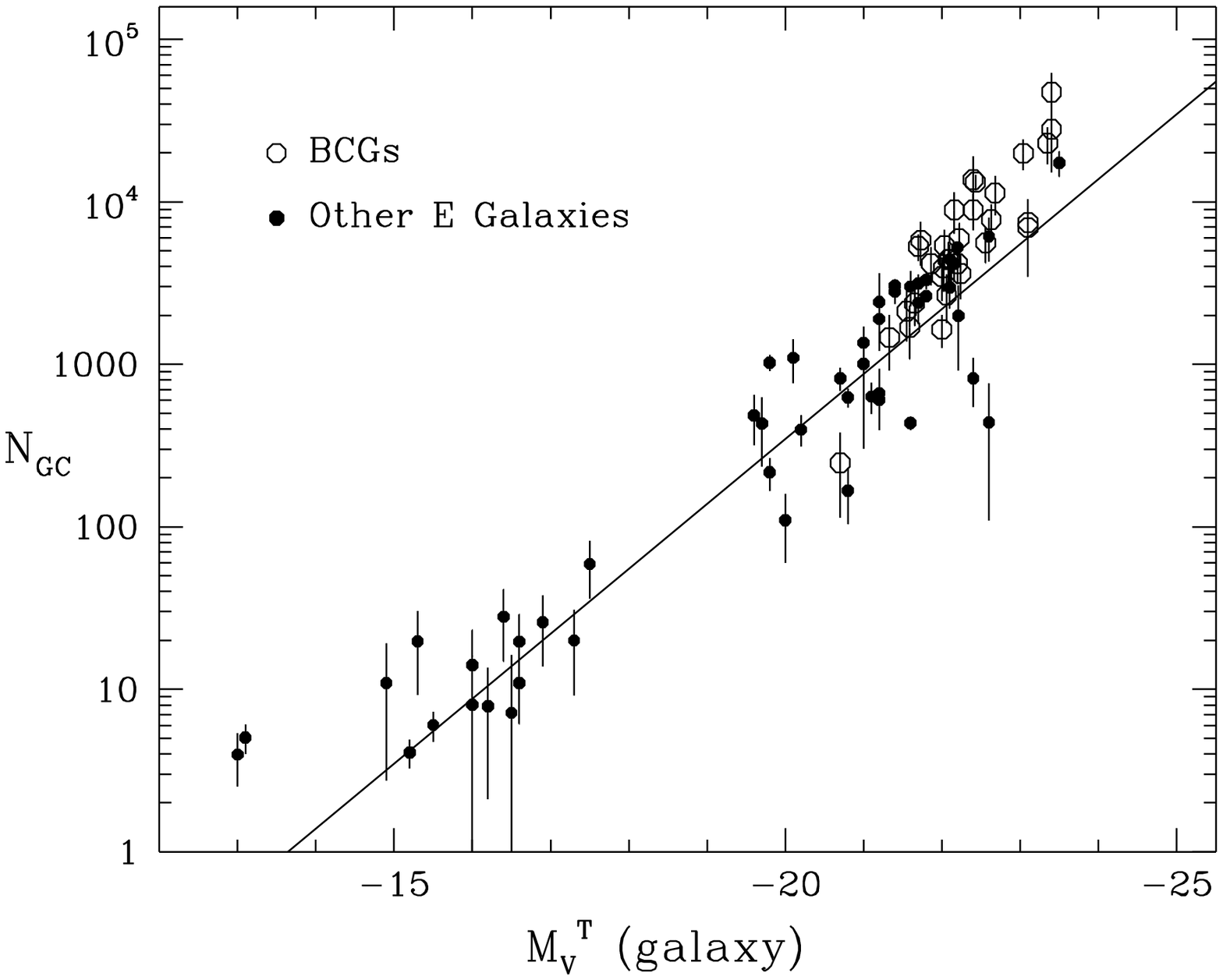]{Total number of globular clusters plotted against
galaxy luminosity, for elliptical galaxies.  {\it Open circles} indicate
giant ellipticals which are the centrally dominant members of their clusters
(BCGs).  {\it Solid dots} indicate all other E-type galaxies, including dwarfs.
The solid line is a direct proportionality $N_{GC} \sim L_{gal}$ for a 
specific frequency $S_N^0 = 3.5$. See Sec.~5.4 of the text.
\label{fig11}}

\figcaption[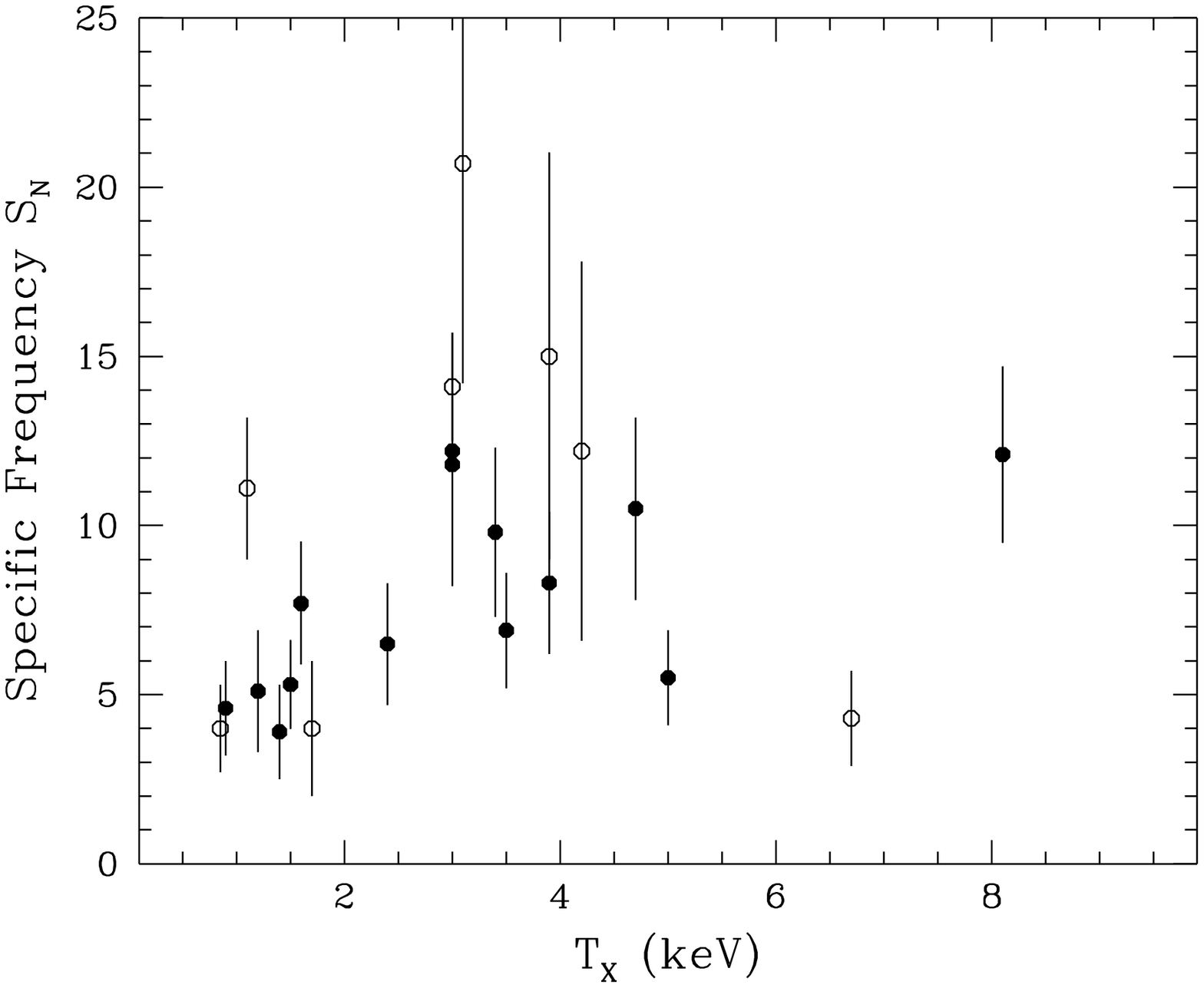]{Specific frequency $S_N$ plotted against the
X-ray temperature of the ICM gas.  Open symbols are BCGs taken from 
Table 6.  Solid symbols are BCGs taken from the
\protect\cite{btm97} sample, where we have converted their metric $S_N^{40}$ to the
global specific frequency via $S_N = 1.3 S_N^{40}$ (see Sec.~5.4).
\label{fig12}}

\figcaption[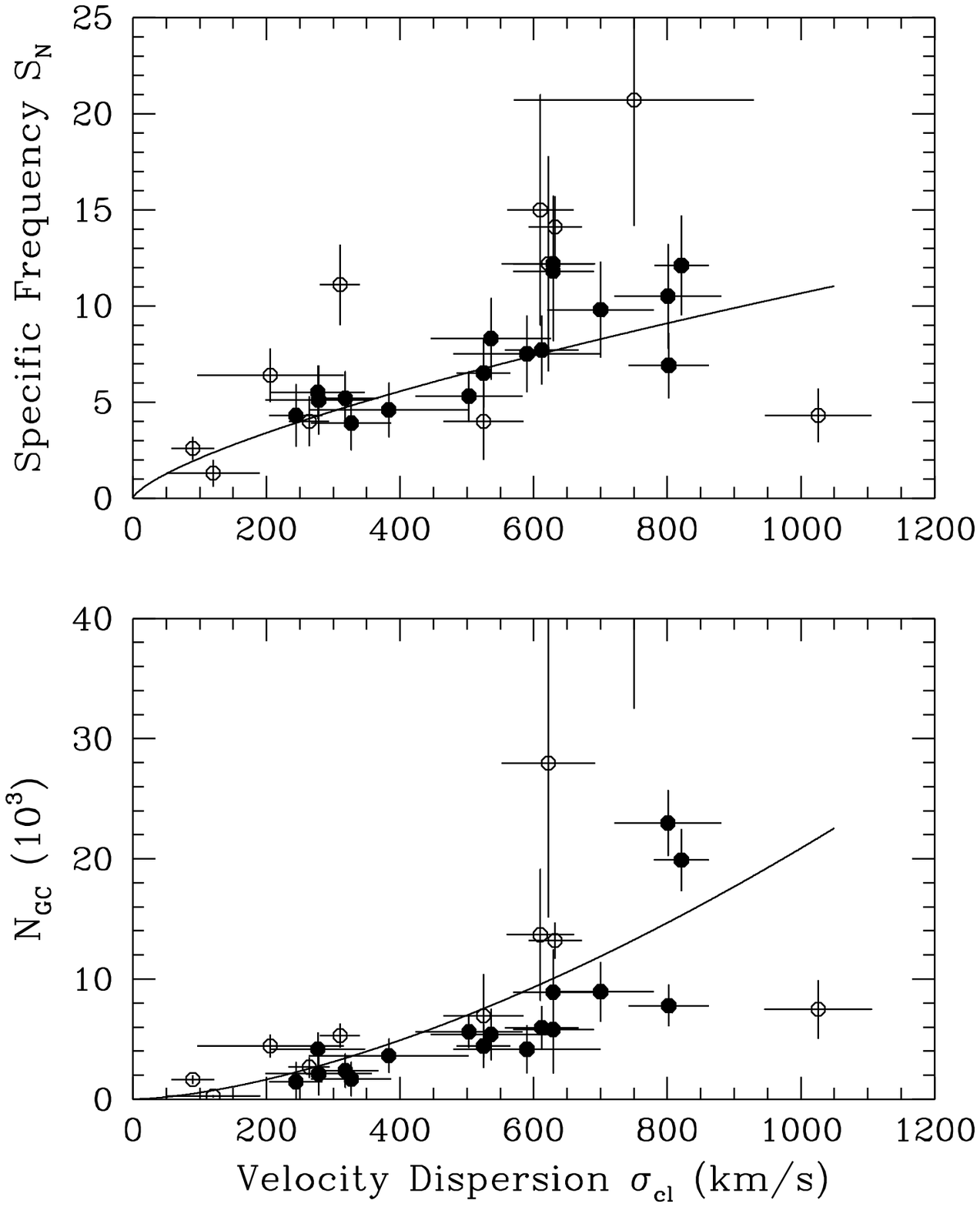]{(a) {\it Upper panel:} BCG specific frequency
plotted against the velocity dispersion $\sigma_{cl}$ of the galaxies in the cluster.
Solid symbols are the BTM97 sample, and open symbols are taken from 
Table 6 as above.  The solid line shows the scaling relation
from Eq.~(6).  (b) {\it Lower panel:}  Total cluster population $N_{GC} \sim L \cdot S_N$
plotted against $\sigma{cl}$.  Symbols are as above.  The solid line shows
the scaling relation from Eq.~(17).
\label{fig13}}

\figcaption[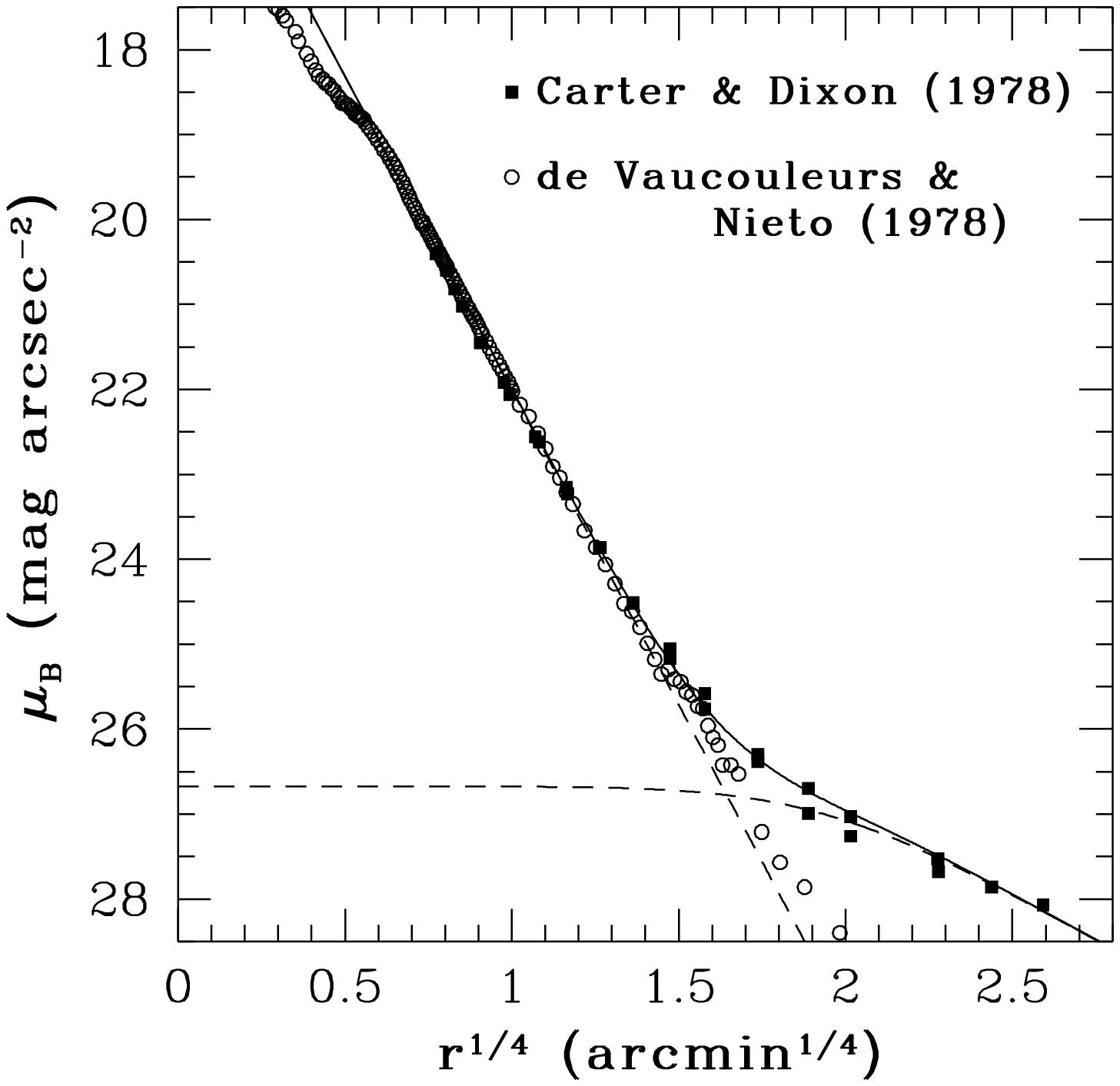]{Radial profile of the M87 halo light, deconvolved
into `body' and `envelope' components as described in Sec.~5.4 of the text.
Solid dots represent the $B-$magnitude surface intensity profile of Carter \& Dixon
(1978); open symbols represent the $B$ measurements of \protect\cite{dvn78}.
The extended envelope component, defined to fit the Carter-Dixon data,
is assumed to be intracluster light in the overall potential well of the Virgo
cluster; see text.
\label{fig14}}

\figcaption[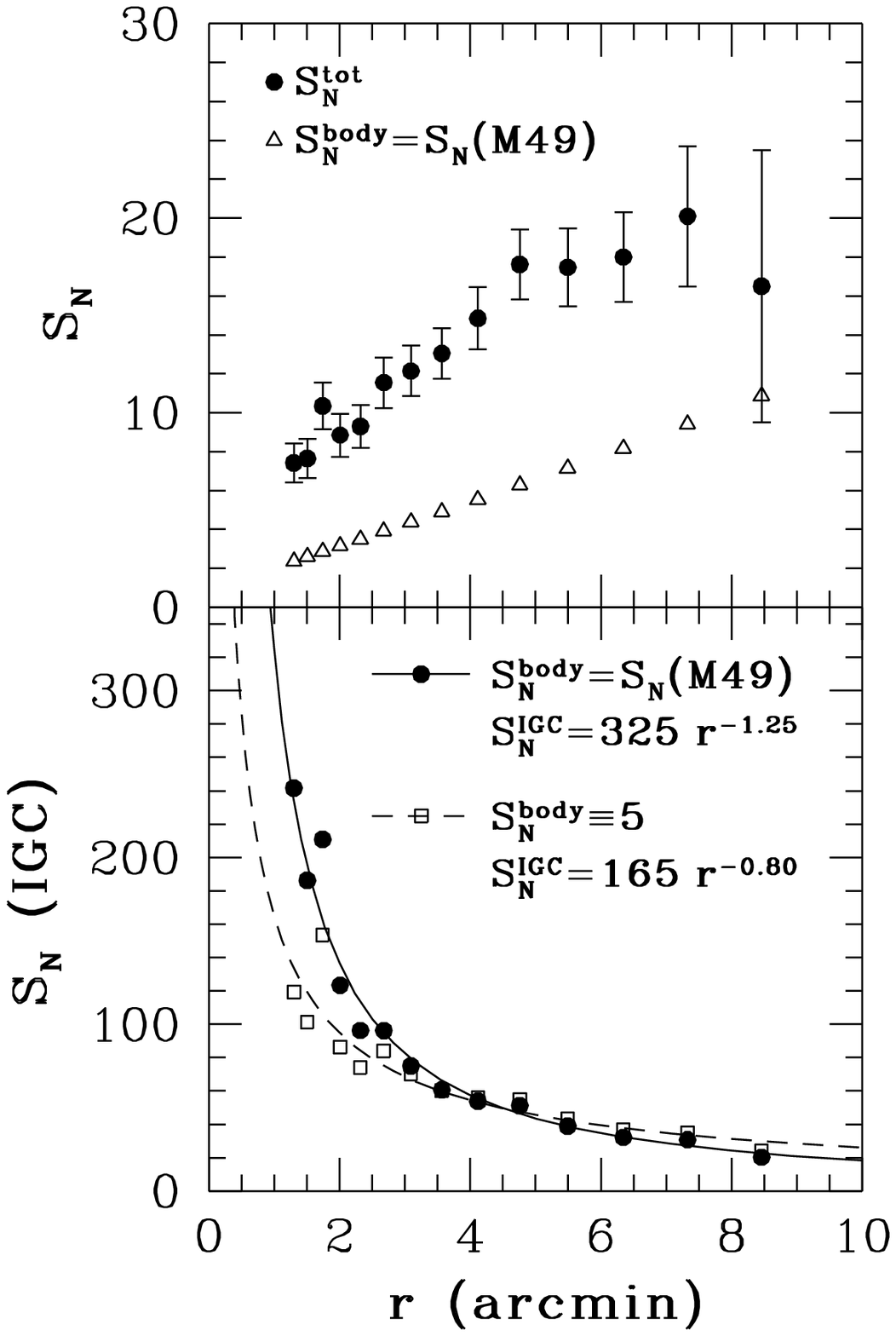]{(a) {\it Upper panel:}  Specific frequency profiles
for the globular cluster systems of M87 (solid symbols) and M49 (open symbols), 
as described in Sec.~5.4 of the text.  A model for the M87 GCS is assumed here in
which the main body of the galaxy has a normal M49-like specific frequency,
while the intergalactic globular clusters from the extended envelope raise the
total specific frequency up to the observed value $S_N^{tot}$.
Both sets of data are 
taken from Fig.~21 of \protect\cite{mhh94}, and have been properly scaled 
to reflect a Virgo distance modulus of $(m-M)_V=31.0$ and the 
same GCLF turnover and dispersion as adopted in Sec.~3.
(b) {\it Lower panel:}  Deduced specific frequency profile for the IGC 
(envelope) component, obtained from the difference between the top and bottom curves
in panel (a).  The assumed IGCs are required to have an extremely high overall
$S_N$, and to be concentrated toward the galaxy nucleus.
\label{fig15}}

\figcaption[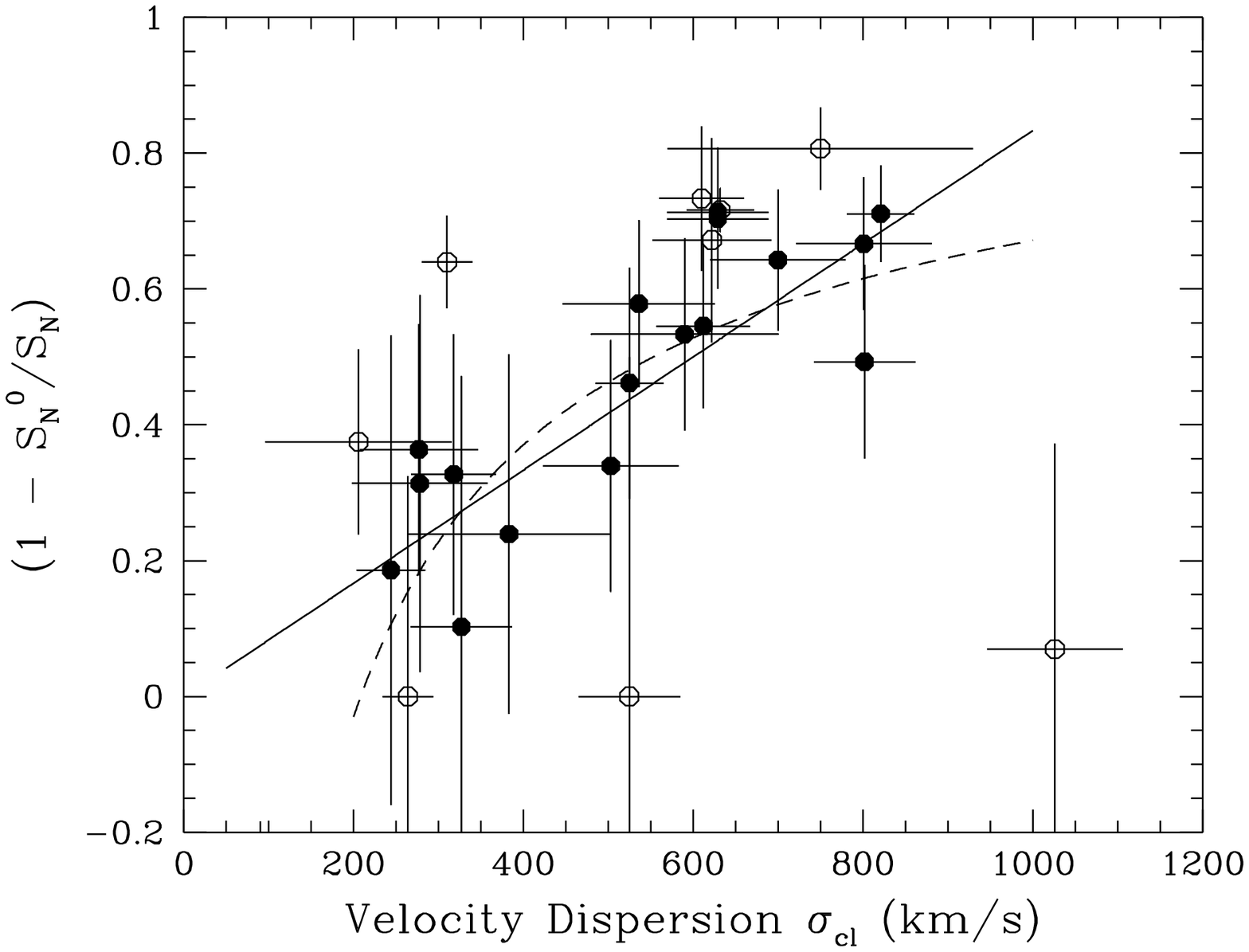]{Mass fraction $f_M = 1 - (S_N^0/S_N)$ plotted against
$\sigma_{cl}$.  According to the galactic wind hypothesis discussed in
Sec.~5.5, $f_M$ is the fraction of the initial protogalactic gas lost to the ICM,
$f_M = M_{gas}/(M_{gas} + M_{stars})$.  The {\it dashed line} indicates the
predicted dependence of $f$ on $\sigma$ from the scaling relations in Eqs.~(7-9),
while the {\it solid line} shows the more approximate linear relation defined
in the text.
\label{fig16}}

\figcaption[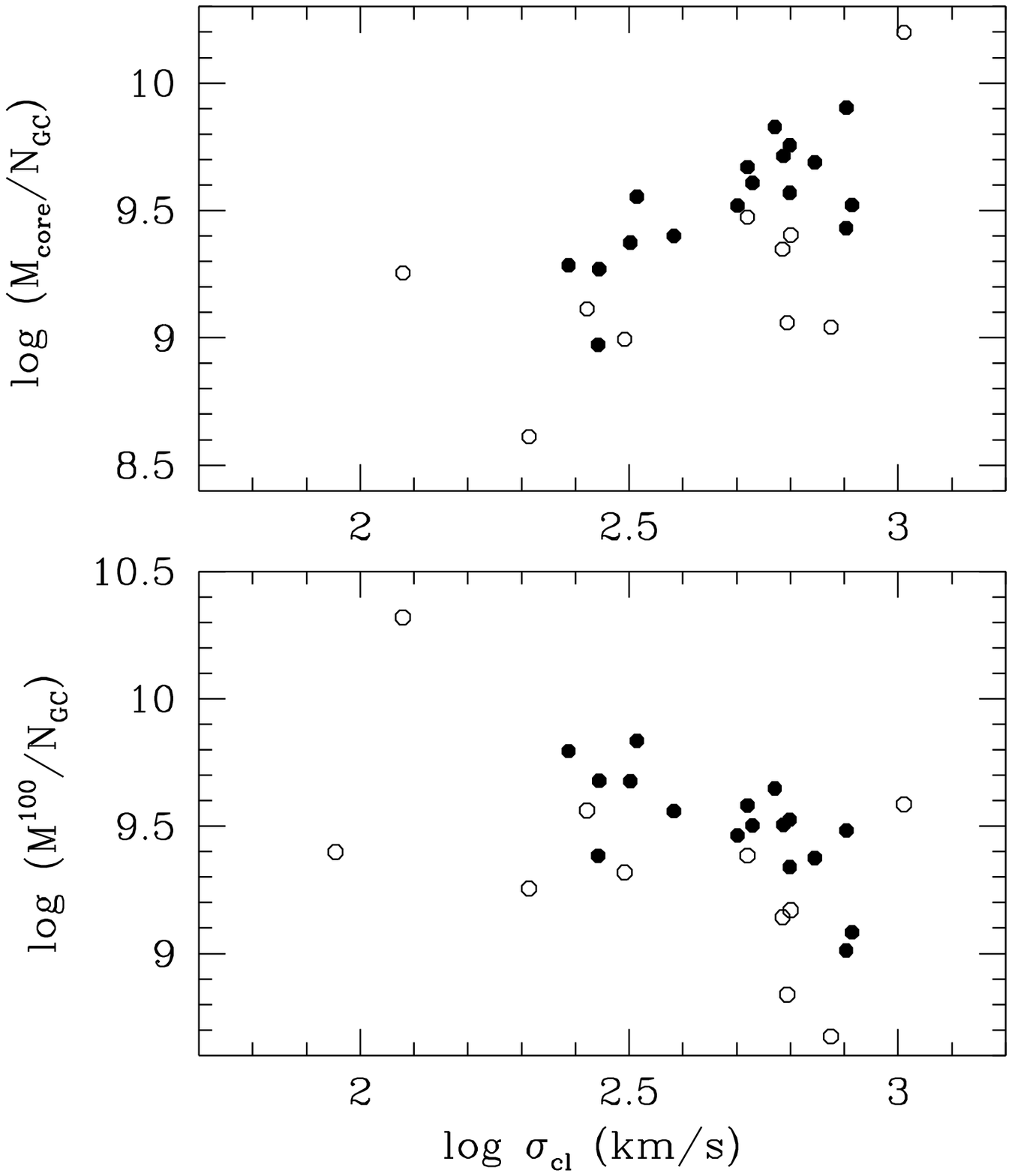]{(a) {\it Upper panel:}  Ratio of total core mass
in the cluster of galaxies to globular cluster population, plotted against
velocity dispersion.  See Sec.~5.5.  (b) {\it Lower panel:}  Ratio of ``metric''
core mass (within 100 kpc radius) to $N_{GC}$, plotted against velocity 
dispersion.
\label{fig17}}

\end{document}